\documentclass{aa}  

\usepackage{graphicx}
\usepackage{txfonts}
\usepackage{arydshln}
\usepackage{xcolor}
\usepackage[normalem]{ulem}
\DeclareUnicodeCharacter{2212}{-}
\DeclareUnicodeCharacter{223C}{~}

\definecolor{mydarkgreen}{RGB}{0,100,0}

\usepackage{multirow} 
\usepackage{hyperref}

\begin{document}

\title{Development of convective envelopes in massive stars}

\subtitle{Implications for gravitational wave sources}

   \author{A. Romagnolo
          \inst{1}\fnmsep\inst{2}
          \and
          J. Klencki\inst{2}\fnmsep\inst{3}
          \and
          A. Vigna-G\'omez\inst{3}
          \and
          K. Belczynski\inst{1}\fnmsep\thanks{Deceased on the 13$^{\rm th}$ of January 2024.}
          }

   \institute{Nicolaus Copernicus Astronomical Centre, the Polish Academy of Sciences, ul. Bartycka 18, 00-716 Warsaw, Poland\\
   \email{amedeoromagnolo@gmail.com}
   \and
   European Southern Observatory, Karl-Schwarzschild-Strasse 2, 85748, Garching bei München, Germany
   \and
   Max Planck Institute for Astrophysics, Karl-Schwarzschild-Strasse 1, 85748, Garching bei München, Germany}
   \date{}

 
  \abstract
   {The structure of stellar envelopes strongly influences the course and outcome of binary mass transfer, in particular of common-envelope (CE) evolution. Convective envelopes can most easily be ejected during CE events, leading to short-period binaries and potentially gravitational-wave (GW) sources. Conversely, radiative envelope are thought to lead to CE mergers and Thorne-\.Zytkow objects (T\.ZOs) or quasi-stars (QS).
   }
  {Rapid binary models based on \cite{Hurley_2000} often assume that any CE event with a Hertzsprung-gap donor results in a CE merger, in tension with literature. We improve this with a more self-consistent criterion based on the presence of a convective envelope.
  } 
   {Using 1D stellar models ({\tt MESA}), we systematically investigate the development of convective envelopes in massive stars. We provide fitting formulae for rapid binary codes and implement them into the {\tt StarTrack} population synthesis code to refine the CE treatment and examine the impact on GW sources, T\.ZOs, and QSs.}
   {We show that convective envelopes in massive stars are highly sensitive to the treatment of superadiabacity and the mixing length. Our revised CE model significantly reduces (factor $\sim$20) the predicted merger rate of binary black hole (BH-BH) mergers with total masses between $\sim$20 and 50~$M_\odot$. This leads to a bimodal mass distribution with a strong metallicity dependence. We also predict that the current T\.ZO/QS formation rate in the Galaxy (up to $\sim$10$^{\rm -4}$~yr$^{\rm -1}$), combined with their predicted lifetimes, makes their detection unlikely.}
  {Our study strongly suggests that the role of CE evolution in the formation of BH-BH mergers has been considerably overestimated for BH-BH mergers with $M_{\rm tot}$\,$\geq$\,20~$M_\odot$. We highlight that any prediction from the CE channel for massive BH-BH mergers (>50~$M_\odot$) heavily hinges on our limited understanding of stellar structure and mass loss close to the Eddington limit.  }


   \maketitle
%

\section{Introduction}
There is growing evidence that most massive stars are born in binaries or higher order multiples \citep{Sana2012,Moe2017}.  When a star expands as a binary component, it may overflow its Roche lobe and start transferring mass onto its companion. During a mass transfer (MT) event the orbit is altered due to the combination of (i) mass exchange between the two binary components and (ii) partial or complete ejection of the donor star’s envelope from the system, which removes angular momentum from the binary.  Such MT events are thought to play a key role in the formation of vast classes of astrophysical objects: Algol systems \citep{Hilditch_2001,Matson_2016}, X-ray binaries \citep{Tauris_2006,van_den_Heuvel_2019}, stripped stars and stripped supernovae \citep{Paczynski_1967,vanbeveren_1998,Gotberg_2018,Shenar_2020b,Drout_2023,Gotberg_2023}, stellar mergers \citep{Schneider_2019,Hirai_2021}, and gravitational wave (GW) sources \citep{deMink_2016,Marchant_2016,Tauris_2017,Belczynski_2020,Mandel_2022}.
The detectable GW sources that can form via MT are close-orbit double compact objects (DCOs) that merge within the age of the Universe: binary black hole (BH-BH), binary neutron star (NS-NS) and black hole-neutron star (BH-NS) systems. Within the first three observing runs the LIGO/Virgo/KAGRA (LVK) collaboration has detected 76 BH-BH, 2 NS-NS and 4 BH-NS \citep{Abbott_2019,Abbott2021,Abbott_2023}. DCOs at short orbital separations are potential candidates for becoming GW sources, since their separation needs to be sufficiently small for them to merge within a Hubble time ($\sim$13.6~Gyr) and potentially be a detectable LVK source. The evolution of isolated binaries through the MT channel is a promising avenue to explain, at least partially, the formation of GW sources due to the role of MT in reducing the orbital distance of massive binaries. Besides the MT isolated binary evolution, several other channels have been proposed, such as chemically homogeneous evolution \citep{deMink_2016,Marchant_2016,Mandel_2016} in isolated binaries on tight orbits, dynamical evolution in globular clusters \citep{Spitzer_1969,Kulkarni_1993,Sigurdsson_1993,PortegiesZwart_2000,Rodriguez_2015,diCarlo_2019}, nuclear clusters \citep{Miller_2009,Antonini_2016}, dynamical evolution through the Kozai-Lidov mechanism \citep{Kozai_1962,Lidov_1962,Thompson_2011,Antonini_2017} within hierarchical triples and multiples \citep{Liu_2018,VignaGomez_2021}, pairing and growth of stellar mass BHs in the disks of active galactic nuclei \citep{McKernan_2014,Stone_2017,Samsing_2022}, and primordial BH binaries \citep{Sigurdsson_1993,Bird2016,Haimoud2017,Chen_2018}. In this paper, we will focus exclusively on the MT scenario in isolated binaries.

The structure of a star that will experience Roche-lobe overflow (RLOF) is crucial to the fate of a MT episode. Generally, stars with radiative(convective) envelopes are more likely to experience stable(unstable) MT episodes. Dynamically unstable MT episodes result in a common envelope (CE) phase \citep{Paczynski1976,Webbink_1984,van_den_Heuvel_2019,Ivanova2020}. The CE phase sees the envelope of the donor star engulf the entire binary system. Because of the resulting drag, the donor core and the accretor spiral inwards on a dynamical timescale, leading to a drastic decrease in orbital separation. If the orbital energy of the binary is enough to gravitationally unbind the envelope, the inspiral stops and the binary becomes a short-period system: an ideal GW source candidate. Otherwise, the inspiral leads to a merger between the two objects. We define this evolutionary pathway, where stars interact through at least one CE event, as the CE channel, while we call the evolutionary pathway in which binaries evolve only through dynamically stable MT events as the stable mass transfer (SMT) channel. 
While a CE event remains a highly complex phenomenon that we cannot fully model \citep{Ivanova2020}, several studies argue that it may only lead to close DCOs if  the CE phase takes place in wide binary orbits where their donor giants have an outer convective envelope \citep{Kruckow_2016,Klencki_2020,Marchant_2021,Klencki_2021}. This is due to the fact that extended convective envelopes, (i) are loosely gravitationally bound (i.e. low binding energy $E_{\rm bind}$) and can be ejected during a CE event, (ii) have a steep density gradient at the core-envelope boundary that sets an end point for the CE inspiral, and (iii) are thought to expand on the adiabatic timescale in response to mass loss, which makes the MT more likely to become unstable \citep{Ivanova_2013,Ge_2015,Pavlovskii_2015,Ge_2020}. 
This implies that interacting binaries with donors with an extended, convective envelope are more likely to initiate a CE event as well as to survive it \citep{Podsiadlowski_2001}. On the other hand, radiative envelopes are centrally condensed, with minimal mass in the outer 10\% of the stellar radius \citep{Podsiadlowski_2001}. If a CE event occurs in a binary with a radiative envelope donor, the high envelope binding energy demands a substantial amount of orbital energy for a successful CE ejection, while the lack of a steep density gradient at the core-envelope boundary makes it difficult for the inspiral to halt. As a result, a CE phase initiated by a radiative-envelope giant is thought to lead to a stellar merger. A possible outcome of a CE episode is that of a merger between a star and a compact-object companion. One possibility is a Thorne–\.Zytkow object (T\.ZO). A T\.ZO is a hypothetical binary merger product with an accreting NS in the core of a non-degenerate star \citep{Thorne_1975,Thorne_1977} that is predicted to manifest as a cool supergiant \citep{Cannon_1992,Farmer_2023}. Depending on the T\.ZO mass, its hydrostatic equilibrium can be sustained either through nuclear burning on or near the NS surface, or via accretion onto the NS \citep{Eich_1989,Biehle_1991}. Another hypothetical object with similar formation channels is what is sometimes referred to in the literature as a quasi-Star (QS), which is a rapidly accreting BH inside a non-degenerate star, where the accretion rate can exceed the Eddington limit  \citep{Begelman_2008,Bellinger_2023}. On top of the T\.ZO formation scenarios, a QS can also form through the entrapment of a primordial BH inside a proto-planetary disk, forming what is known as a Hawking star \citep{Hawking_1971,Bellinger_2023}. 

In this study we employ the 1D stellar evolution code \textsc{Modules for Experiments in Stellar Astrophysics} \citep[{\tt MESA};][]{Paxton2011, Paxton2013, Paxton2015, Paxton2018, Paxton2019} to determine the transition point at which massive stars develop outer convective envelopes. Using our detailed stellar models, we develop analytical fits, which are then implemented within the {\tt StarTrack} rapid binary evolution code \citep{Belczynski_2002,Belczynski_2008}. Our objective is to quantify the implication of the development of convective envelopes on the outcome of CE phases within isolated binary populations – a key channel for forming GW sources and T\.ZOs/QSs. We explore two primary uncertainties: the radial expansion of massive stars and the treatment of superadiabaticity in radiation-dominated regions of stellar envelopes. We then discuss the statistical and physical properties of the population of GW sources and T\.ZOs/QSs, as well as the key uncertainties.

\section{Methods} \label{sec:method}

Here we describe the adopted 1D stellar evolution models used to study the development of outer convective envelopes (Section~\ref{subsubsec:conv_env}). Subsequently, we introduce the setup of our rapid population synthesis calculations, which we use to compute DCO mergers and T\.ZOs/QSs arising from isolated binary evolution. We then explain the implementation within our population synthesis models of the analytical fits for outer convective envelope development, which we use for a preliminary determination of the outcome of CE episodes. Finally, we discuss how our models affect the formation rate and detectability of T\.ZOs/QSs.

\subsection{1D stellar evolution: {\tt MESA}}
\label{subsubsec:conv_env}

We use {\tt MESA} to develop analytical fits for the presence of an outer convective envelope, with application for rapid binary evolution codes. These fits establish a threshold effective temperature $T_{\rm eff}$, below which stars develop an outer convective envelope as a function of metallicity and luminosity. All our models consider non-rotating stars.


\paragraph{\underline{Development of outer convective envelopes}}
Following \cite{Klencki_2020,Klencki_2021}, we define outer convective envelopes (from now simply addressed as "convective envelopes") as stellar envelopes that have convective near-surface regions that make up at least 10\% of their total envelope mass. As an example, we show in Appendix~\ref{sec:outconvenv} the evolution of a 50~$M_\odot$ star at both 0.1~$Z_\odot$ and $Z_\odot$ until its convective envelope reaches 70\% of the total envelope mass, following the models described below. We highlight that the 10\% criterion is a low threshold to choose, where a few percent of the envelope mass can be in sub-surface convective zones around the opacity peaks. Additionally, the 10\% convective region near the surface gets stripped quickly during MT. By making this choice we are being optimistic in how many stars will we treat as convective-envelope donors in terms of CE ejection.

\begin{table*}[h!]
    \centering
    \caption{Adopted models with respective assumptions for the determination of the CE outcome prior to the $\alpha\lambda$ calculation, mixing length and density inversion treatment (with or without \textit{MLT++}).
    }
    \resizebox{\linewidth}{!}{
    \begin{tabular}{l|ccc|cc}
    \hline
         Model & Preliminary CE & $\alpha_{\rm ML}$ & \textit{MLT++} & References & Comments on\\
          & outcome determination & & & & stellar models \\  
         \hline
         M0 (default) & Evolutionary type & 2.0 & No & \cite{Belczynski_2002,Belczynski_2008,Belczynski_2010} & From \cite{Pols_1998}\\
         M\_$\alpha_{\rm ML}$1.5 & Envelope type & 1.5 & No & \cite{Klencki_2020} & --\\
         M\_$\alpha_{\rm ML}$1.82\_MLTpp & Envelope type & 1.82 & Yes & This study & \textit{MLT++} only for $Z$\,$\geq$\,0.5~$Z_\odot$\\
    \hline
    \end{tabular}}
    \label{tab:models}
\end{table*}

\paragraph{\underline{M\_$\alpha_{\rm ML}$1.5}} The first model is derived from the fits provided by \cite{Klencki_2020}, which come from {\tt MESA} simulations made with semiconvection factor $\alpha_{sc}$ = 100, step-overshooting of 0.345 for the convective core, mixing length $\alpha_{\rm ML}$ of 1.5, and the Ledoux criterion to determine the convective boundaries \citep{Ledoux_1947}. We will refer to this model as M\_$\alpha_{\rm ML}$1.5.

\paragraph{\underline{M\_$\alpha_{\rm ML}$1.82\_MLTpp}}
We developed a grid of models where we employ the Ledoux criterion with step-overshooting of 0.5 \citep{Scott_2021} for the hydrogen-burning convective core, exponential overshooting of 0.01 for the other convective regions, and the Dutch wind prescription from \cite{Glebbeek_2009} based on \cite{Vink2001,deJager_1988,NugisLamers_2000}. Considering that $\alpha_{\rm ML}$\,=\,1.5 does not reproduce the estimated radii of RSGs \citep{Chun_2018}, we set $\alpha_{\rm ML}$\,=\,1.82 \citep{Choi_2016} for this model. We apply these initial conditions to our simulations for two different metallicity values: Z\,=\,0.0142 ($\sim$$Z_\odot$; \citealt{Asplund_2009}) and Z\,=\,0.00142 ($\sim$0.1~$Z_\odot$). For the sake of numerical stability for $Z_\odot$ stars we use the \textit{MLT++} \citep{Paxton2013} method to reduce superadibaticity in regions near the Eddington limit (see Section~\ref{subsec:uncert} for more details). Hot wind-driven mass loss rates are calculated with the \cite{Vink2001} model for homogeneous optically-thin winds. With the discovery of clumped winds behavior in OBA-type stars, the current estimates for mass loss rates were shown to be 2-3 times lower \citep{Bouret_2005,Surlan_2012,Surlan_2013,Sander_2017,GormazMatamala_2021} than what is currently considered as the default in many stellar evolution models. In order to mitigate a potential overestimation of mass loss rates at high metallicities we therefore calibrate the $Z_\odot$ simulations by scaling down the wind mass loss by half. This is accomplished by setting the \texttt{Dutch\_scaling\_factor} to 0.5\footnote{We acknowledge that reducing the \textit{Dutch\_scaling\_factor} in {\tt MESA} not only reduces the mass loss rates for OBA-type stars, but also the ones for both the dust-driven winds \citep{deJager_1988} and optically thick winds \citep{NugisLamers_2000}. However, a reduction in the mass loss rates for these winds might be justified by the recent literature (wind-driven mass loss rates comparisons: for dust-driven winds see Section~3 of \citealt{Decin_2024} and Section~5 of \cite{Antoniadis_2024}; for thick and thin winds see Section~2 of \citealt{GormazMatamala_2024} and Appendix~A of \citealt{Romagnolo2024})}. With these models we evolved stars until the end of their core-helium burning phase.

\subsection{Rapid population synthesis: {\tt StarTrack}}
\label{subsec:ST}

In our study we use the {\tt StarTrack}\footnote{\url{https://startrackworks.camk.edu.pl}} rapid population synthesis code \citep{Belczynski_2002,Belczynski_2008} to simulate the formation and mergers of DCOs and T\.ZOs/QSs. The code adopts the evolutionary fitting formulae from \cite{Hurley_2000}, based on the stellar models by \cite{Pols_1998}, in order to emulate the evolution of single and binary stars for a wide array of initial conditions and physical assumptions. The currently implemented stellar physics, star formation history, metallicity (we set solar metallicity at $Z_\odot$\,=\,0.0142) as well as detectability criteria for GW sources in \cite{Belczynski_2020}, with two modifications described in Section~2 of \cite{Olejak_2020}.

\paragraph{\underline{Initial Conditions}}
The primary star mass (M1) is drawn from the broken power-law initial mass function from \cite{Kroupa_1993,Kroupa_2002} between 5 and 150 $M_\odot$, while the secondary mass star is derived from the mass of the primary times the mass ratio factor ($q$) from a uniform distribution \textit{q} $\in$ [0.08/M1; 1]. To produce the semi-major axis of the binary system we used the initial orbital period ($P$) power law distribution f(log($P$/d))\,$\sim$\,log($P$/d)$^{\rm \alpha_P}$, with log($P$/d)\,$\in$\,[0.5; 5.5] and $\alpha_P$\,=\,$-$0.55, and a power law initial eccentricity distribution with an exponent $\alpha_e$\,=\,$-$0.42 within the range [0; 0.9].These distributions are based on \cite{Sana2012} and extended from log$P$\,=\,3.5 up to log$P$\,=\,5.5, following \cite{deMink_2015}. 

\paragraph{\underline{Wind-driven Mass Loss and Accretion}}
The stellar winds prescription adopted in \textsc{StarTrack} is described in \cite{Belczynski_2008} with the Luminous Blue Variable (LBV) mass loss from \cite{Belczynski_2010}. .
\paragraph{\underline{Compact Object Remnants}}
We adopt the weak pulsation pair-instability supernovae and pair-instability supernovae formulation from \cite{Belczynski_2016,Woosley_2017}, which limits the mass spectrum of single BHs to 50 $M_\odot$. We also use a delayed supernova engine \citep{Fryer_2012,Belczynski_2012}, which affects the birth mass of NSs and BHs so that it allows for the formation of compact objects (COs) within the first mass gap ($\sim$2\,-\,5~$M_\odot$), with the assumption that anything beyond 2.5 $M_\odot$ is a BH and therefore every non-white dwarf degenerate object under that mass is a NS \citep{horvath2020}. Additionally, we use a Maxwellian distribution of natal kicks of $\sigma=265$ km s$^{-1}$ during core-collapse, damped by the amount of non-ejected mass, represented by a fallback factor \citep{Fryer_2012}. 

\paragraph{\underline{Mass Transfer}}
The adopted prescription for the accretion onto a CO both during a SMT event or from stellar winds is based on the approximations from \cite{King2001}. A 50\% conservative SMT is used for non-degenerate accretors ($f_a$ = 0.5) and the remaining mass (1 - $f_a$) being dispersed in the binary surroundings with a consequent loss of orbital momentum modeled following \cite{Belczynski_2008}. For the stability criterion during MT (and therefore initiation of a CE phase) we refer to \cite{Belczynski_2008}.

\subsubsection{Common Envelope}
\label{subsubsec:CE}

\paragraph{\underline{Preliminary determination of CE outcome}}

Prior to any estimate of the envelope binding energy and binary orbital energy, {\tt StarTrack} uses the donor's evolutionary type (based on \citealt{Hurley_2000} nomenclature) to determine whether the binary will merge inside the CE or evolve trough the $\alpha\lambda$ formalism.
If the donor is a star in its MS, HG or Wolf-Rayet (WR) phase, the system is always assumed to merge during a CE phase \citep{Belczynski_2008}. The assumption that HG donors lead to a CE merger was motivated by the fact that a strong density gradient is needed at the core-envelope boundary to stop the inspiral. It was suggested that the CHeB phase may be a good proxy for such a gradient, with HG donors instead having a more continuous density distribution in which the dynamical inspiral would never stop \citep{Belczynski_2010}. However, the definitions of both core-envelope boundary and HG phase are unclear \citep[e.g.][]{Deloye_2010,Marchant_2021}, and it has been argued that the HG vs. CHeB distinction in the \cite{Hurley_2000} types is not a good proxy for the presence of a density drop \citep{Kruckow_2016,Klencki_2020,Klencki_2021,Marchant_2021}. The steep core-envelope density gradient, as well as envelope binding energies sufficiently low for unbinding, are instead a characteristic feature of stars with deep convective envelopes \citep[e.g.,][]{Dewi_2000,Tauris_2001,VignaGomez_2022}.
Stars with radiative envelopes are generally less evolved and more centrally condensed, unlike stars with convective envelopes. This makes the ejection of convective envelopes possible according to the $\alpha\lambda$ formalism \citep{Kruckow_2016,Klencki_2020,Klencki_2021,Marchant_2021}. We therefore use the presence of a convective envelope criterion as a preliminary check to quickly determine the fate of CE events prior to any calculations for CE evolution.

\paragraph{\underline{CE evolution}}
We use the $\alpha\lambda$ energy formalism \citep{Webbink_1984,LivioSoker_1988,deKool_1990} to calculate the outcome of CE events initiated by convective donors. We adopt 100\% efficiency of the orbital energy transferred ($\alpha_{\rm CE}$=1) into the envelope. The binding energy model comes from \cite{Dominik_2012}, where the $\lambda_{\rm CE}$ parameter was fit to include internal energy in the energy budget and meant to be an average between the $\lambda_{\rm b}$ and $\lambda_{\rm g}$ prescriptions presented in \cite{XuLi_2010,Xu_2010}. In successful CE events, the entire envelope of the donor star is assumed to be ejected from the binary. The accretion during CE events is zero with stellar accretors and 5\% the Bondi rate with CO accretors \citep{MacLeod_2017}.

\subsubsection{Convective Envelope Threshold in {\tt StarTrack}}
\label{subsubsec:conv_env_ST}

In our models for the determination of the outcome of CE events, we do not check for the evolutionary stage of the giant donor, as it is for our reference model that we address as M0. We only check whether, according to the fits derived from the  {\tt MESA} tracks described in Section~\ref{subsubsec:conv_env}, the donor's effective temperature ($T_{\rm eff}$) is below the threshold for the development of a convective envelope. If the donor star has a convective envelope, the binary evolves through the $\alpha\lambda$ \cite{Webbink_1984,LivioSoker_1988} formalism. If not, the outcome of the CE episode is that the binary system merges. In M\_$\alpha_{\rm ML}$1.82\_MLTpp, only two metallicities were considered from the {\tt MESA} simulations. For $Z$\,<\,0.5~$Z_\odot$ ($Z$\,$\geq$\,0.5~$Z_\odot$), we use the fits from the {\tt MESA} tracks at $Z$\,=\,0.1~$Z_\odot$ ($Z_\odot$ with reduced wind mass loss rates and \textit{MLT++}).

\subsubsection{Summary of the {\tt StarTrack} models}

We list below all the models that we adopt in {\tt StarTrack} to calculate the physical properties of binary populations (Table~\ref{tab:models}), where we investigate different assumptions for the development of convective envelopes and the preliminary determination of the outcome of CE events.

\paragraph{\underline{M0}} Our reference model.  This model is the only one where we determine the outcome of a CE event via the evolutionary type of the donor star rather than the envelope type. This means that prior to the CE evolution calculations through the $\alpha\lambda$ formalism, any massive binary at the onset of a CE phase with a donor star in its MS, WR or HG phase will not expel the CE and face a merger event\footnote{As described in \cite{Belczynski_2008}, the standard prescription in {\tt StarTrack} for the presence of convective envelopes in massive giants is set to $T_{\rm eff}\leq$~5370~K.}.

\paragraph{\underline{M\_$\alpha_{\rm ML}$1.5}} We adopt the fits from \cite{Klencki_2020} for the development of convective envelopes. If at the onset of a CE envelope the donor star has a radiative envelope, it will lead the binary to a merger. Otherwise, the evolution of the binary will follow the $\alpha\lambda$ formalism.

\paragraph{\underline{M\_$\alpha_{\rm ML}$1.82\_MLTpp}} We use the fits from the {\tt MESA} models described in Section~\ref{subsubsec:conv_env} to determine the presence of a donor star with a convective envelope.

\paragraph{\underline{RMAX submodels}} The stellar tracks from \cite{Hurley_2000} may be overestimating radial expansion of massive stars \citep{Grafener_2012,Agrawal_2022,Agrawal_2022b,Romagnolo_2023}. To circumvent this, we explore model variations in which the expansion of stars in {\tt StarTrack} is limited, guided by detailed stellar models (see \citealt{Romagnolo_2023} and Appendix~\ref{sec:RMAX} for details). We therefore introduce submodels M0$_{\rm RMAX}$, M\_$\alpha_{\rm ML}$1.5$_{\rm RMAX}$ and M\_$\alpha_{\rm ML}$1.82\_MLTpp$_{\rm RMAX}$.

\subsection{Population estimates of Thorne-\.Zytkow objects and quasi-stars in the Galactic disk}
\label{subsubsec:formations}

It is still unclear whether a NS or a stellar-mass BH can sink down to the center of a massive star without tidally disrupting the core first \citep[e.g.,][]{Metzger_2022,HutchinsonSmith_2023,Everson_2024}. In this study we assume that in every merger outcome of a CE episode with a CO accretor results in the assembly of a T\.ZO/QS. 

\paragraph{\underline{T\.ZOs lifetimes}}
The lifetime of T\.ZOs was shown by \cite{Farmer_2023} to be proportional to their mass. However, their parameter space only covered masses between 5 and 20~$M_\odot$. We address this by linearly interpolating lifetimes within this range based on reported values for $\tau_{\rm min}$ ($M_{\rm TZO}\sim$\,5~$M_\odot$) and $\tau_{\rm max}$\,($M_{\rm TZO}\sim$\,20~$M_\odot$). For higher and lower masses, we extrapolate these values. To account for uncertainties in red supergiant (RSG) pulsations, we explore two scenarios (i) Weak RSG pulsation scenario, where weak pulsation-driven mass loss allows for the T\.ZOs to live longer ($\tau_{\rm min}$\,=\,5\,$\times$\,$10^4$~yr; $\tau_{\rm max}$\,=\,2\,$\times$\,$10^5$~yr; $\tau_{\rm TZO}$\,=\,$10^4M_{\rm TZO}$~yr), and (ii) Strong RSG pulsation scenario (following \citealt{Yoon_2010} mass loss rates and implications from \citealt{Farmer_2023}), where extreme mass loss rates due to strong RSG pulsations lead to lower lifetimes ($\tau_{\rm min}$\,=\,$10^2$~yr; $\tau_{\rm max}$\,=\,$10^3$~yr; $\tau_{\rm TZO}$\,=\,6$\times10^1M_{\rm TZO}-2\times10^2$~yr). For instance, \cite{Nathaniel_2024} use a different prescription for $\tau_{\rm TZO}$ (see Appendix~\ref{sec:lifetimes} for a comparison between their $\tau_{\rm TZO}$ model and ours). They fit the weak RSG pulsation lifetimes from the \cite{Farmer_2023} results and assume flat $\tau_{\rm TZO}$ values outside the 5-20~$M_\odot$ range. In contrast, we linearly interpolate between $\tau_{\rm min}$ and $\tau_{\rm max}$, followed by extrapolation for higher and lower T\.ZO masses.

\paragraph{\underline{QSs lifetimes}}
\cite{Bellinger_2023} showed that the lifetime of QSs $\tau_{\rm QS}$ might be correlated, at least for MS stars with accreting BHs in their center, to the mass ratio $M/M_{\rm BH}$ between the non-degenerate component and the BH. 
Their study focused on a wide array of QSs, also referred to as Hawking stars, arising from primordial BHs of even substellar masses, where the $M/M_{\rm BH}$ ratio can be considerably high. In contrast, our study focuses on QSs formed from stellar-mass BHs. Here we therefore adopt a simplified approach, assuming the lifetime of these QSs depends primarily on their total mass. We utilize the same mass dependence described for T\.ZOs under the Weak and Strong Red Supergiant (RSG) pulsation scenarios to estimate their lifetimes.

\paragraph{\underline{Galactic disk model}} 
The disk has an estimated total mass of (5.17\,$\pm$\,1.11$)\times10^{10}$\,$M_\odot$ \citep{Licquia_2015}. We make the same assumptions for initial conditions that were discussed in \cite{Olejak_2020b}: the thin disk evolution, which accounts for 90\% of the disk mass, is approximated by ten star formation episodes, which started 10 Gyr ago and lasted until the present time. Each episode lasts 1 Gyr, with a constant star formation rate (SFR) of $\approx$5~$M_\odot$\,yr$^{-1}$. Using the metallicity-age relation from \cite{Haywood_2013}, the formation episodes start at an initial metallicity of $Z$\,=\,0.1~$Z_\odot$ 10 Gyr ago, which increases by 0.1\,$Z_\odot$ at each consecutive formation episode until it reaches $Z_\odot$ during the last 1~Gyr.
Since in both the  Strong and Weak RSG Pulsation scenarios T\.ZOs/QSs have a lifetime that is orders of magnitude less than 1\,Gyr, we can assume that none of them can come from the thick disk, since the contribution of stars younger than 9~Gyr in this region can be considered negligible \citep{Cariulo_2004,Cignoni_2006}. The only star formation episode that is of interest to us is the most recent one at $Z_\odot$ in the thin disk. All the initial conditions are the same of the GW population simulations described in Section~\ref{subsec:ST}. We rescale a synthetic population of 400,000 binaries to the stellar mass of the last 1~Gyr formation history in the thin disk and randomize the formation age along the look-back time (uniform distribution) of each binary to check if each of our synthetic T\.ZOs/QSs (i) had enough time to form and (ii) is young enough to still be present in the Galaxy. 

\section{Results and Discussion}
\label{sec:Results}

\subsection{M\_$\alpha_{\rm ML}$1.82\_MLTpp: Convective envelope boundary}
\label{subsec:DCE_bound}

\begin{figure}[ht]
    \centering
    \includegraphics[width=0.495\textwidth]{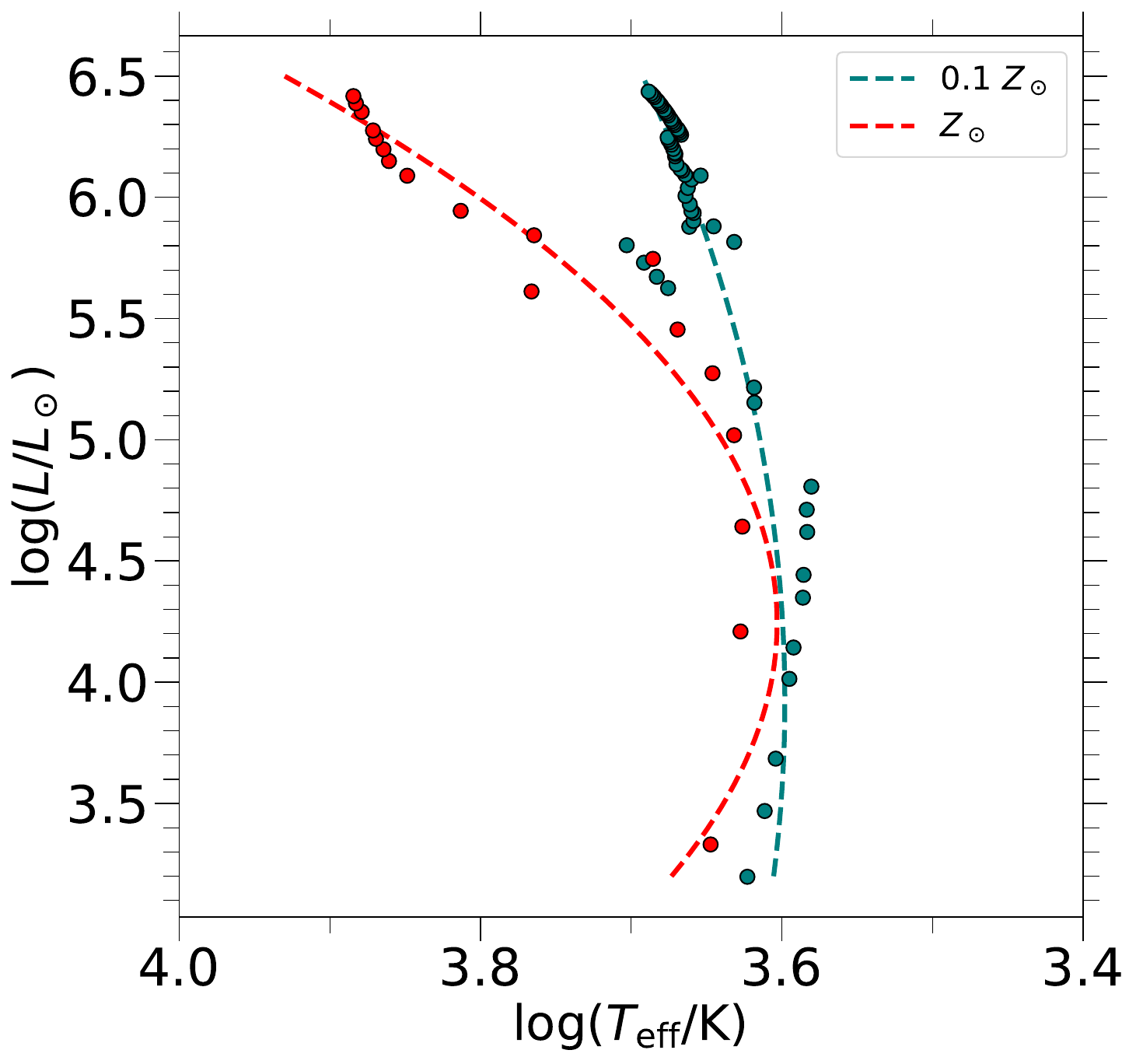}
    \caption{HR diagram with the positions of stars from our {\tt MESA} models where they develop convective envelopes for metallicities $Z$\,=\,0.1\,$Z_\odot$ (teal) and $Z$\,=\,$Z_\odot$ (red). Their respective fits (Equations~\ref{eq:conv_0.1Zsun} and \ref{eq:conv_Zsun}) are represented by the dashed lines.
    }
    \label{fig:conv_fit}
\end{figure}

Figure~\ref{fig:conv_fit} shows a Hertzsprung–Russell (HR) diagram with the position of stars from our {\tt MESA} simulations (Section~\ref{subsubsec:conv_env}) at which they first develop convective envelopes. We then fit their HR position to establish a threshold effective temperature $T_{\rm eff}$ (in units of K) as a function of stellar luminosity (in units of $L_\odot$) at which massive stars develop convective envelopes. For 0.1~$Z_\odot$, we fit the following threshold

\begin{equation}
    {\rm log}T_{\rm eff} = 1.422\times10^{-2}({\rm log}L)^2 -1.117\times10^{-1}{\rm log}L + 3.817,
    \label{eq:conv_0.1Zsun}
\end{equation}

and for our $Z_\odot$ models

\begin{equation}
    {\rm log}T_{\rm eff} = 6.427\times10^{-2} ({\rm log}L)^2 -5.456\times10^{-1}{\rm log}L+4.761.
    \label{eq:conv_Zsun}
\end{equation}

In Figure~\ref{fig:conv_fit} we compare these fits (Equations~\ref{eq:conv_0.1Zsun} and \ref{eq:conv_Zsun}) with the {\tt MESA} models from which they were developed. For the fit from the $Z_\odot$ tracks we get a $\chi^2$ value of 0.0031, while for the ones at 0.1~$Z_\odot$ we get $\chi^2$\,=\,0.0034 .

\cite{Picker_2024} recently proposed a method to identify stars that have developed convective envelopes, assuming a constant $T_{\rm eff}$ boundary for all masses and luminosities at a given metallicity. This method was adopted for a grid of {\tt MESA} models with $M_{\rm ZAMS}$ up to 30~$M_\odot$ with the assumption of no wind-driven mass loss. In contrast, our study examines how $T_{\rm eff}$ depends on stellar luminosity via two models adopting different assumptions for internal mixing, wind mass loss, and superadiabaticity, and $M_{\rm ZAMS}$ up to 100~$M_\odot$. Our models reveal a significant dependence of the $T_{\rm eff}$ boundary for convective envelope formation on stellar luminosity (see also Figure~6 from \citealt{Klencki_2020}, which brought similar results), in particular for tracks where density inversion suppression was enforced via \textit{MLT++} (details in Section~\ref{subsec:conv_pop}). Our assumptions allow us to develop criteria for convective envelope formation that account for the influence of stellar luminosity without the need to extrapolate for $M_{\rm ZAMS}$ up to 100~$M_\odot$.

\subsection{Yields of the CE channel}
\label{subsec:conv_pop}

\begin{figure*}[h!]
\centering
\includegraphics[width=0.497\linewidth]{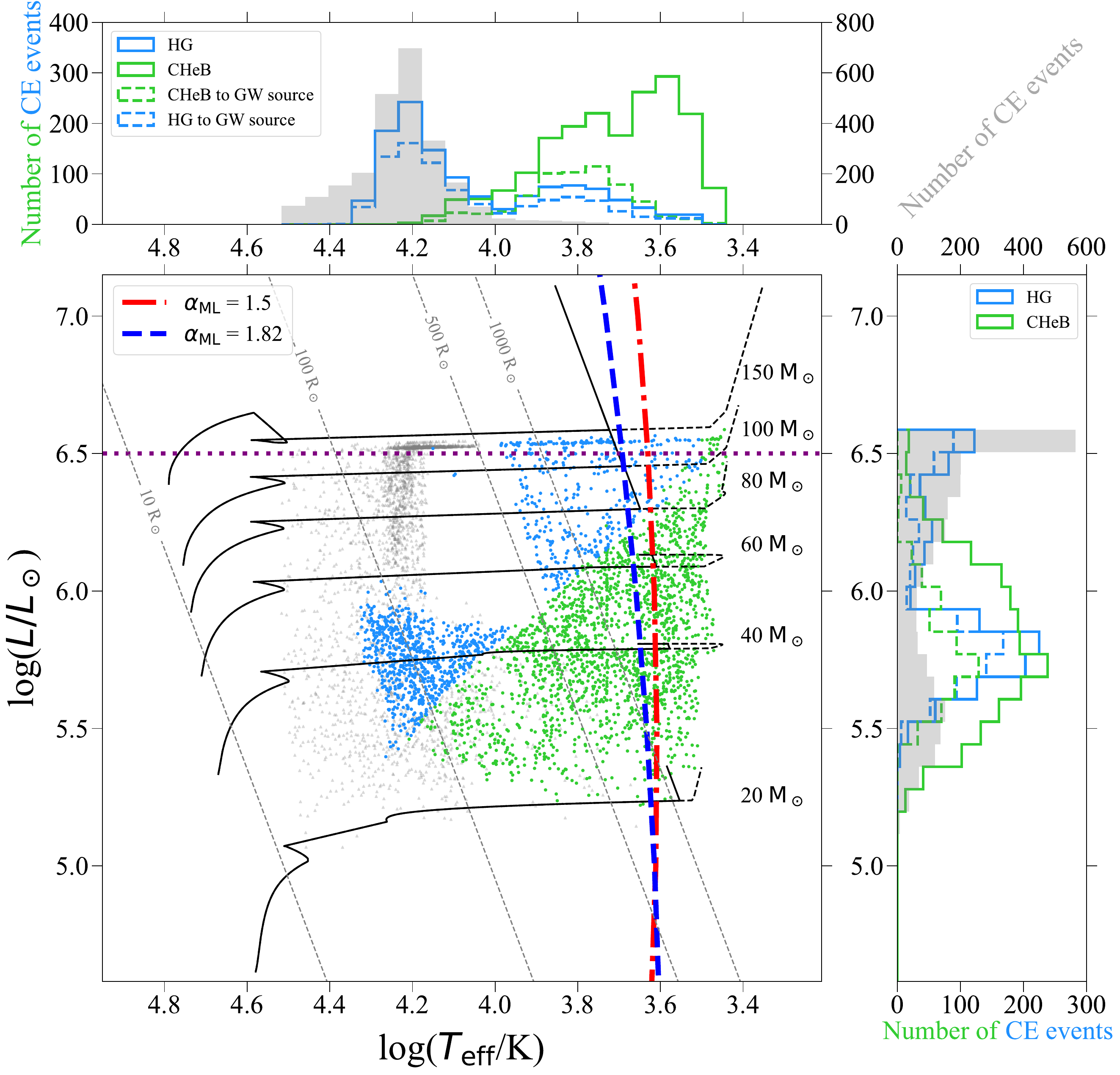}  
\includegraphics[width=0.497
    \linewidth]{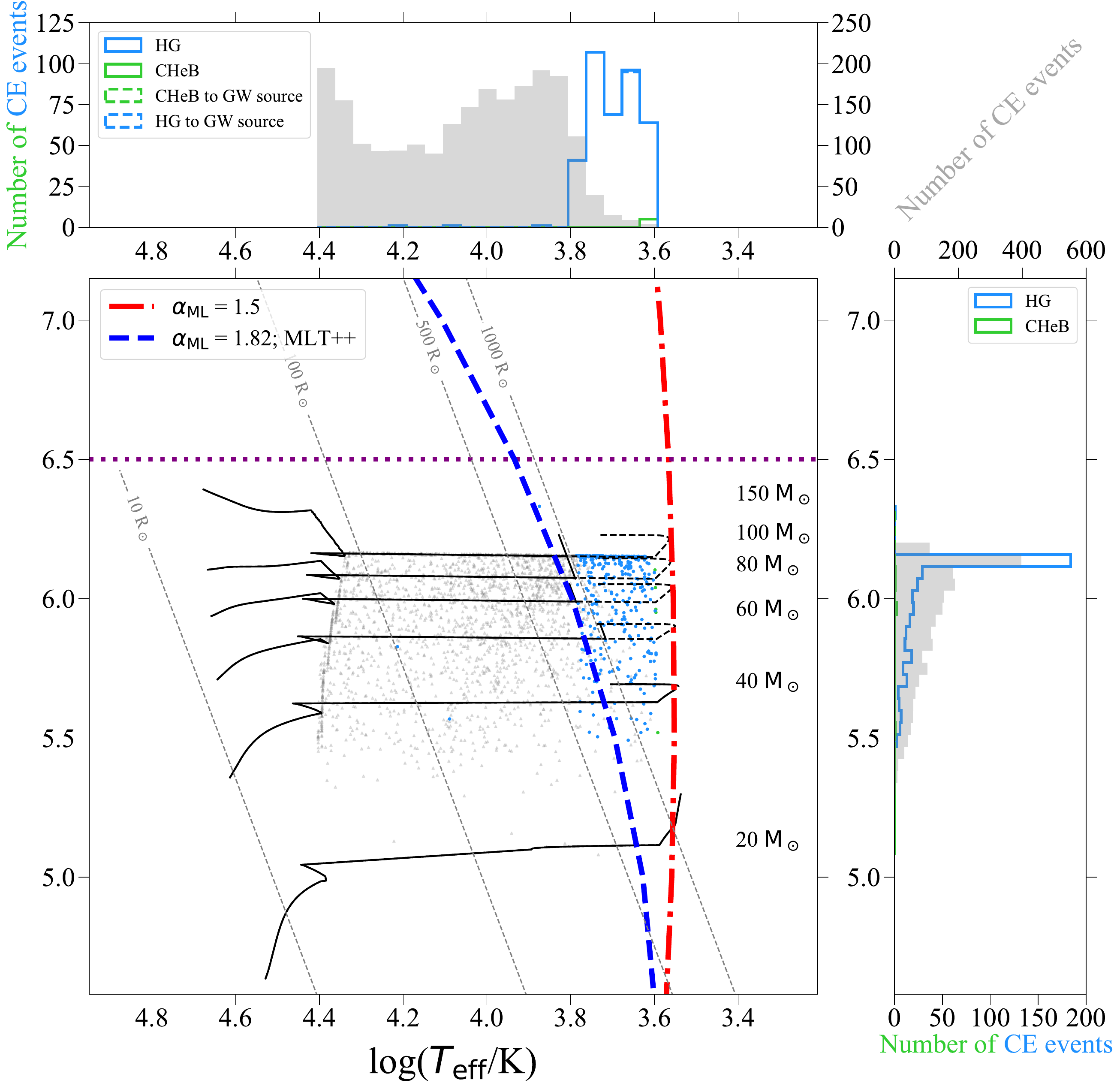}  

  \caption{HR diagram position of the donor stars at the onset of a CE event with a BH accretor at $Z$\,=\,0.1~\,$Z_\odot$ (left) and $Z_\odot$ (right), with six evolutionary tracks for single stars at $M_{\rm ZAMS}$\,=\,20, 40, 60, 80, 100, and 150~$M_\odot$ for M0 (black dashed line) and M0$_{\rm RMAX}$ (black solid line). The gray dots represent the CE events with a BH companion that lead to a stellar merger and the potential formation of a QS, while the blue and green dots represent respectively HG and CHeB stars (i.e. evolutionary types in {\tt StarTrack}, following \citealt{Hurley_2000}) in binaries that successfully expel the CE and become BH-BH. The red and blue lines represent the threshold for respectively M\_$\alpha_{\rm ML}$1.5 and M\_$\alpha_{\rm ML}$1.82\_MLTpp where a star becomes cold enough to develop a convective envelope. The horizontal purple dotted line represents the luminosity threshold beyond which we extrapolate the convective boundary tracks. We plot the distributions of $\log T_{\rm eff}$ and $\log L$ values of donors that have led to the CE mergers (in gray) or to CE ejections and the formation of BH-BH systems (in green or blue, depending on the donor evolutionary type in {\tt StarTrack}). We highlight that in the right panel there is a near-complete overlap between the BH-BH progenitor binaries and the ones that become GW sources. Dashed-line histograms correspond only to BH-BH systems that merge within a Hubble time.
  }
  \label{fig:HR_conv}
\end{figure*}

\paragraph{\underline{Evolutionary stage vs. envelope type}} 
Figure \ref{fig:HR_conv} shows two HR diagrams with the log$T_{\rm eff}$ boundaries from M\_$\alpha_{\rm ML}$1.82\_MLTpp and M\_$\alpha_{\rm ML}$1.5 . In each plot we show the HR position of the CE donors with a BH accretor from two synthetic populations of 400,000 binaries each, respectively at $Z$\,=\,0.1~$Z_\odot$ and $Z$\,=\,$Z_\odot$. We plot HG and CHeB donors within binaries that evolve past the CE phase and become BH-BH determined solely based on the $\alpha\lambda$ prescription. HG and CHeB donors are shown in the plots with blue and green dots. We clarify that our presentation of the evolutionary stages of massive stars follows from the nomenclature of stellar types introduced by \cite{Hurley_2000}. We make this distinction because in our default M0 model it is assumed that HG donors do not lead to CE ejection, and only the stars depicted in green in Figure~\ref{fig:HR_conv} can lead their binary to evolve into BH-BH systems through the CE channel. In Models M\_$\alpha_{\rm ML}$1.5 and M\_$\alpha_{\rm ML}$1.82\_MLTpp, instead of distinguishing between HG and CHeB stellar types, we use the envelope type to determine the outcome of CE events. Only donors on the right-hand (the cooler) side of the dashed lines are allowed to survive the CE phase and potentially become GW sources.

\paragraph{\underline{CE donors at $Z$\,=\,0.1~$Z_\odot$}} 
According to the convective envelope boundary defined in Equation~\ref{eq:conv_0.1Zsun}, roughly 82\% and 87\% (respectively, from M\_$\alpha_{\rm ML}$1.82\_MLTpp and M\_$\alpha_{\rm ML}$1.5) of the BH-BH mergers from model M0 originate from systems that initiate a CE phase with a radiative envelope donor (left panel of Figure~\ref{fig:HR_conv}).  In models M\_$\alpha_{\rm ML}$1.5 and M\_$\alpha_{\rm ML}$1.82\_MLTpp those systems are predicted to instead merge during the CE phase and produce a T\.ZO/QS. On the other hand, we note that above $\sim$80~$M_\odot$ the \cite{Hurley_2000} HG donors can develop convective envelopes.  This effect reshapes the predicted BH-BH total mass distribution (Section~\ref{subsec:GW_properties} for more details) and contributes towards a higher yield of GW sources from the CE channel with respect to the M0 model. The difference between M\_$\alpha_{\rm ML}$1.5 and M\_$\alpha_{\rm ML}$1.82\_MLTpp $T_{\rm eff}$ thresholds is due to the different assumptions for $\alpha_{\rm ML}$, with $\alpha_{\rm ML}$\,=\,1.82 finding better support in the derived $T_{\rm eff}$ of RSGs \citep{Chun_2018}. The higher the $\alpha_{\rm ML}$ parameter, the wider the convective regions, which means a more massive convective envelope. This in turn means that stellar envelopes reach the threshold for 10\% of their mass to be convective at higher $T_{\rm eff}$ values, i.e. earlier in their evolution.

\paragraph{\underline{CE donors at $Z$\,=\,$Z_\odot$}} 
Compared to the $Z$\,=\,0.1~$Z_\odot$ population, fewer CE events are initiated with a BH accretor at this metallicity, regardless of whether the donor is a HG or CHeB star. This is particularly evident in the right-hand histograms of Figure~\ref{fig:HR_conv}. For log$L$\,<\,6.0, within our synthetic population, 132 and 1091 CE events leading to BH-BH mergers occur respectively at $Z_\odot$ and 0.1~$Z_\odot$. Beyond log$L$\,=\,6.0, instead, 252 at $Z_\odot$ and 346 at 0.1~$Z_\odot$ of CE events are initiated with a significant contribution of highly luminous and cool HG donors. In M0 these donors merge with their BH accretor within the CE due to the criterion based on evolutionary type. At 0.1~$Z_\odot$  many such donors have radiative envelopes and therefore end up as CE mergers also in the updated models. In contrast, at $Z_\odot$ and with \textit{MLT++},
these donors develop convective envelopes. \textit{MLT++} mitigates superadiabaticity in stellar envelopes nearing the Eddington L/M limit, which is particularly important for (very) massive stars as the L/M ratio increases with mass (more details in Section~\ref{subsec:uncert}). This method reduces the temperature gradient \citep{Klencki_2020,Agrawal_2022}, resulting in hotter, smaller supergiants. The inclusion of such luminous HG donors at $Z_\odot$ plays an important role when comparing the merger yield and BH-BH merger properties between different models (see Sections~\ref{subsec:GW_properties} and \ref{subsec:BHBH_MRD_M_Z}). 

\paragraph{\underline{Effect of limiting radial expansion}}
\cite{Romagnolo_2023} showed that interpolating \cite{Hurley_2000} evolutionary tracks at $\sim$0.1~$Z_\odot$ can overestimate radial expansion up to one order of magnitude if compared with 1D stellar evolution models. In M\_$\alpha_{\rm ML}$1.5$_{\rm RMAX}$ and M\_$\alpha_{\rm ML}$1.82\_MLTpp$_{\rm RMAX}$ we circumvent this by restricting the maximum radial expansion of massive stars in {\tt StarTrack}. This in turn prevents some of the wide binaries in our population from interacting in CE events. This limitation affects mainly those stars that develop a convective envelope, since they can only develop it, in terms of $T_{\rm eff}$, relatively near their maximum expansion. This is noticeable in Figure~\ref{fig:HR_conv}, where for $\log L$\,<\,5.7 the bifurcation point between M0 (dotted lines) and M0$_{\rm RMAX}$ (full lines) is close to the $T_{\rm eff}$ thresholds for convective envelope development (red and blue lines). At $\log L$\,$\gtrsim$\,5.7 and $Z$\,=\,$Z_\odot$, instead, M\_$\alpha_{\rm ML}$1.5 shows that the bifurcation point is at considerably higher $T_{\rm eff}$ values than what is needed for stars to develop convective envelope. This means that the initiation of CE events that are ejected by their binaries is predicted to occur in the parameter space where the degree of stellar expansion is rather uncertain.

\subsection{Masses, spins, and merger rates of GW sources}
\label{subsec:GW_properties}

In the local Universe (redshift $z \sim$ 0.2) the LVK collaboration reports within a 90\% credibility interval a merger rate density of 17.9-44 Gpc$^{-3}$yr$^{-1}$ for BH-BH, 7.8-140 Gpc$^{−3}$yr$^{−1}$ for BH-NS and 10-1700 Gpc$^{−3}$yr$^{-1}$ for NS-NS \citep{Abbott_2019,Abbott2021,Abbott_2023}. For BH-BH mergers, the mass distribution was shown to have substructures that cannot fit a single power law, due to two local peaks at $\sim$10 and 35~$M_\odot$ \citep{Abbott2021}. \citep{Abbott2021}inferred the BH-BH effective spin distribution to have a mean centered at $\sim$0.06.

The contribution of the CE channel to the observed population of GW sources is still uncertain \citep{mandel2021rates,Belczynski2022}. 
With the revised treatment of CE evolution many binaries that evolve into GW sources in M0 do not expel the CE and instead merge inside of it. In the following paragraphs we will explore how this assumption changes the estimated merger rate densities within redshift $z$\,<\,0.2 (Table~\ref{tab:MRD}).

\begin{table}[h!]
\centering
\caption{Double compact object local merger rate densities [Gpc$^{-3} $yr$^{-1}$]}
\label{tab:MRD}
\resizebox{\linewidth}{!}{
\begin{tabular}{lccc}
\hline
Model & BH-BH & BH-NS & NS-NS \\
LVK & 17.9 - 44 & 7.8 - 140 & 10 - 1700 \\
\hline
M0 & 65.9 & 13.2 & 112.8 \\
M0$_{\rm RMAX}$ & 63.6 & 11.9 & 114.1\\
\hdashline
M\_$\alpha_{\rm ML}1.5$ & 12.3 & 12.4 & 5.1 \\
M\_$\alpha_{\rm ML}1.5_{\rm RMAX}$& 3.0 & 12.6 & 5.1 \\
\hdashline
\textcolor{red}{M\_$\alpha_{\rm ML}1.82$\_MLTpp} & \textcolor{red}{104.4} & \textcolor{red}{12.8} & \textcolor{red}{17.2} \\
\boldmath
\textbf{\textcolor{mydarkgreen}{M\_$\alpha_{\rm ML}1.82$\_MLTpp$_{\rm RMAX}$}} & \textbf{\textcolor{mydarkgreen}{38.9}} & \textbf{\textcolor{mydarkgreen}{12.5}} & \textbf{\textcolor{mydarkgreen}{15.7}}
\unboldmath
\\
\hline
\end{tabular}}
\tablefoot{Comparison between the LVK local ($z\lesssim$~0.2) merger rate densities \citep{Abbott_2019,Abbott2021,Abbott_2023} and the ones calculated with our models. 
\\
The reference models that base the CE survival on the \cite{Hurley_2000} evolutionary types for post-MS donor stars rather than on the presence of a convective envelope  are the ones with the nomenclature "M0" (the first two from the top).
\\
The model(s) whose maximum stellar expansion and convective envelope prescriptions come from the same set of {\tt MESA} tracks is shown in {\color{mydarkgreen} green}. In {\color{red} red} the most optimistic model for CE survival for BH-BH progenitors. In \textbf{bold} the model(s) that fit(s) within the 90\% credibility interval of the LVK observations.
}
\end{table}

\paragraph{\underline{M\_$\alpha_{\rm ML}$1.5 local merger rates}}
We find a significant reduction in the merger rates of BH-BH and NS-NS systems, respectively by a factor of $\sim$5 (22 in the RMAX variation) and 22. This stems from the fact that in the prior models, most of the CE events leading to GW sources would be initiated by radiative-envelope donors, particularly at low metallicity (Figure~\ref{fig:HR_conv}). With the revised treatment of CE evolution, these events are instead predicted to result in CE mergers and T$\dot {\rm Z}$O/QS formation, leading to a considerable reduction in the number of GW sources. 
The further decline of the BH-BH merger rate in the RMAX variation by an additional factor of $\sim$4 is due to the fact that in massive stars an outer convective envelope only develops when the star is already expanded close to its maximum radius. How much a star expands further from that point is uncertain, particularly in the regime of stars above 40~$M_{\odot}$. In the RMAX variation, the radial expansion along the convective branch in partly quenched (mostly by cool supergiant winds), decreasing the parameter space for CE evolution and GW-source formation. This effect is mass dependent, showing significant impact on the rate of BH-BH mergers but little effect on NS-NS and BH-NS source formation.

\paragraph{\underline{M\_$\alpha_{\rm ML}$1.82\_MLTpp local merger rates}}
In the variation with the assumed mixing length of $\alpha_{
\rm ML} = 1.82$, stars develop convective envelope earlier, at smaller radii (and higher $T_{\rm eff}$). This enlarges the parameter space for CE evolution from convective donors and leads to a smaller decrease in the merger rate of NS-NS and BH-BH sources compared to the M\_$\alpha_{\rm ML}$1.5 model: by a factor of $\sim$7 and $\sim$1.5 (in the RMAX variation), respectively. Surprisingly, we discover a slight increase in the rate of BH-BH mergers in the non-RMAX variation, up to $\sim$100~${\rm Gpc^{-3}yr^{-1}}$. This effect is due to fact that in the revised CE treatment, we model CE evolution from any convective donor, no matter their evolutionary stage. In the past models, CE events initiated by HG giants (according to evolutionary types in the stellar track fits by \citealt{Hurley_2000}) were assumed to always lead to CE mergers. Here, we find that many such stars in the $\gtrsim$\,40~$M_{\odot}$ are in fact convective-envelope giants and therefore ideal candidates for the CE channel (see Figure~\ref{fig:HR_conv}). As a result, in all the updated models we find an increase in the predicted merger rate of the most massive BH-BH mergers (see Section~\ref{subsec:conv_pop} for details). In the M\_$\alpha_{\rm ML}$1.82\_MLTpp specifically, this increase is so significant that it leads to the net increase in the local BH-BH merger rate when all the masses are considered. This is because in this model, the $T_{\rm eff}$ threshold for the formation of a convective envelope in stars above $\sim$\,30\,$M_{\odot}$ shifts to higher $T_{\rm eff}$ values (and smaller stellar radii, enlarging to parameter space for CE evolution) due to the inclusion of \textit{MLT++} in the {\tt MESA} tracks computed at $Z_{\odot}$ metallicity. As a result, the CE channel yield increases significantly for high-metallicity progenitors. As we discuss below, this has an impact for the predictions of how the merger rate evolves with redshift.

\begin{figure}
\centering
\includegraphics[width=1\linewidth]{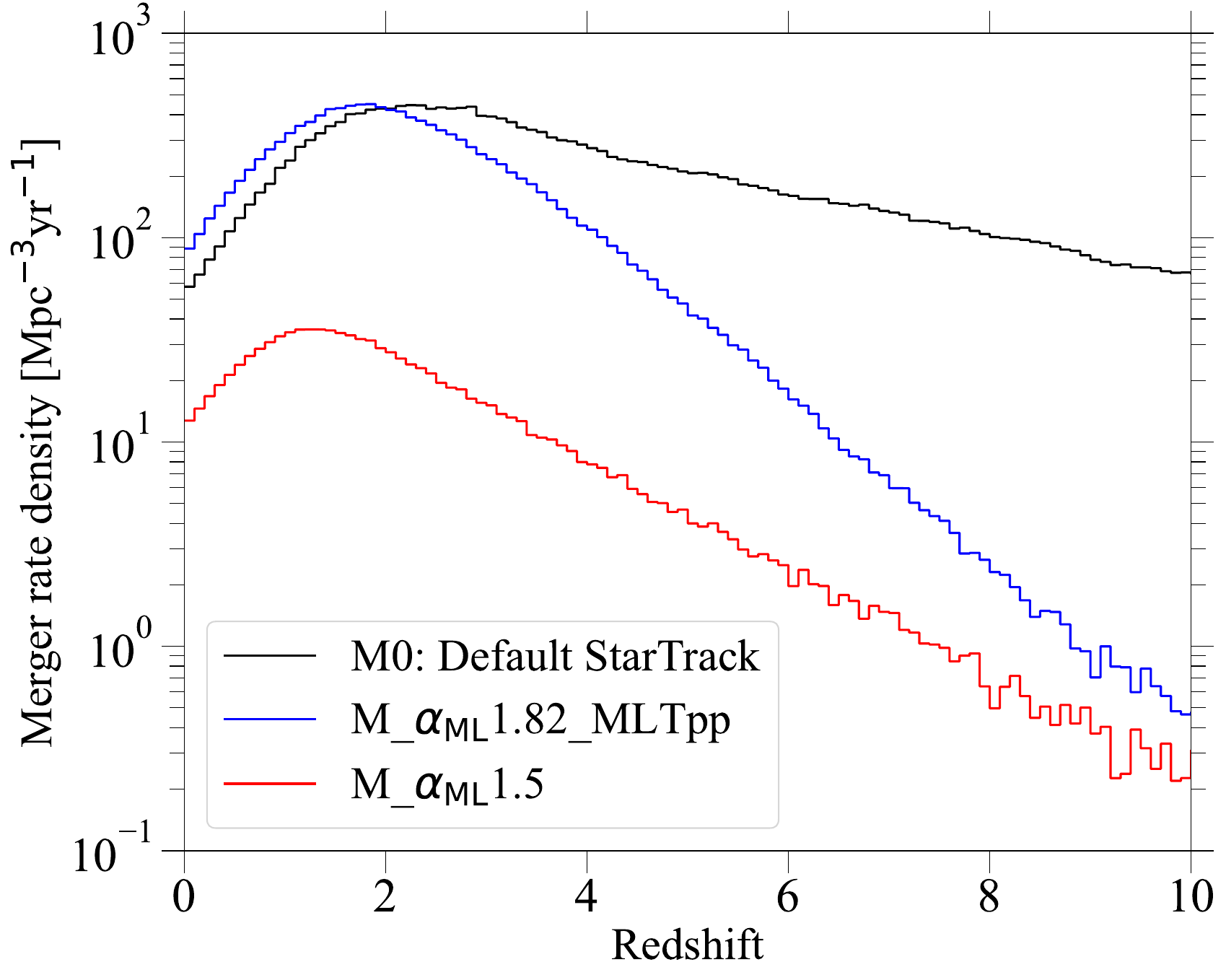} 
    \caption{BH-BH merger rate distribution as a function of redshift. The influence of more optimistic assumptions for convective envelope formation at high metallicities for M\_$\alpha_{\rm ML}$1.82\_MLTpp noticeably increases the mergers rates for the local Universe.}
\label{fig:bhbh_rz}
\end{figure}

\paragraph{\underline{BH-BH merger rate evolution with redshift}}
Figure~\ref{fig:bhbh_rz} shows the merger rate evolution with redshift, revealing a significant variation between models. M0 shows between \textit{z}\,=\,2 and \textit{z}\,=\,10 a slope of log(merger rate density)\,$\approx$\,$-$0.10, while the slopes for M\_$\alpha_{\rm ML}$1.5 and M\_$\alpha_{\rm ML}$1.82\_MLTpp within the same redshift range are respectively roughly $-$0.25 and $-$0.36. The differences are due to changes in the yield of the CE channel for different metallicities. As discussed above, in the M\_$\alpha_{\rm ML}$1.82\_MLTpp model the parameter space for CE ejection is particularly large at high metallicities, guided by the MESA tracks computed at $Z_{\odot}$. As a result, most of the BH-BH mergers in M\_$\alpha_{\rm ML}$1.82\_MLTpp originate from stars formed with $Z$\,>\,0.5~$Z_{\odot}$, which are more common at low redshifts. This leads to an increase in the merger rate density at $z<2$ compared to the M0 model and a steeper decrease in the rate at $z>2$. In M\_$\alpha_{\rm ML}$1.5, on the other hand, the net effect of the revised CE criterion is similar for all the metallicities, with a slightly smaller rate decrease relative to M0 at high $Z$ (due to the inclusion of convective HG donors). 

\begin{figure*}
    \centering
    \includegraphics[width=0.8\textwidth]{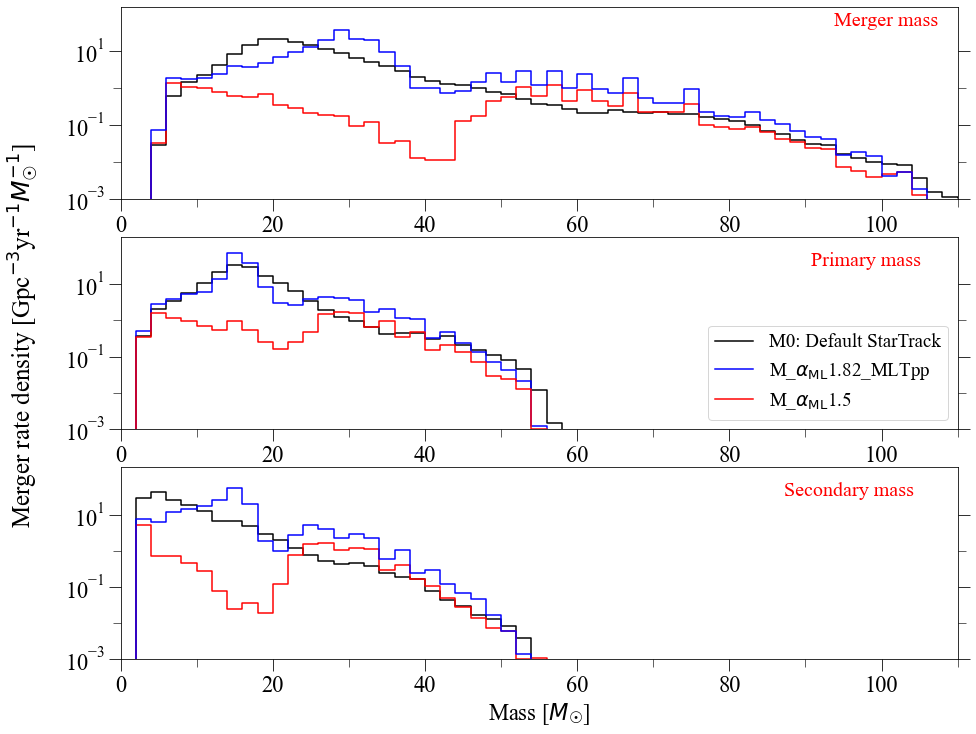}
    \caption{Mass distribution of the BH-BH mergers within redshift 2. In the different panels we show the total mass (top) as well as the mass of the most (middle) and least (bottom) massive BH in the binary.}
    \label{fig:BHBH_mass}
\end{figure*}

\paragraph{\underline{Mass Distribution}}
\label{sec.mass_distr}
Figure \ref{fig:BHBH_mass} shows the distribution within redshift 2 of BH-BH merger, primary and secondary BH mass.
Both M\_$\alpha_{\rm ML}$1.82\_MLTpp and M\_$\alpha_{\rm ML}$1.5 show a bimodal distribution of total masses that M0 does not present and is more akin to LVK observations \citep{Abbott_2019,Abbott2021,Abbott_2023}. M\_$\alpha_{\rm ML}$1.5 brings to a reduction of BH-BH mergers up to a total mass of $\sim$50~$M_\odot$. Beyond that mass, convective-envelope donors are predominantly of the HG type, leading to an increase in the number of CE events. On the other hand, M\_$\alpha_{\rm ML}$1.82\_MLTpp predicts a distribution of masses for BH-BH mergers that peaks around 30 and 55~$M_\odot$. The former total mass is remarkably close to the reported mass of Gaia~BH3 \citep{GaiaBH3}. This coincidental result raises a question on whether BH systems such as Gaia BH3 can be a product of two merging BHs or two close BHs in an inner binary.

\begin{figure}
    \centering
    \includegraphics[width=0.475\textwidth]{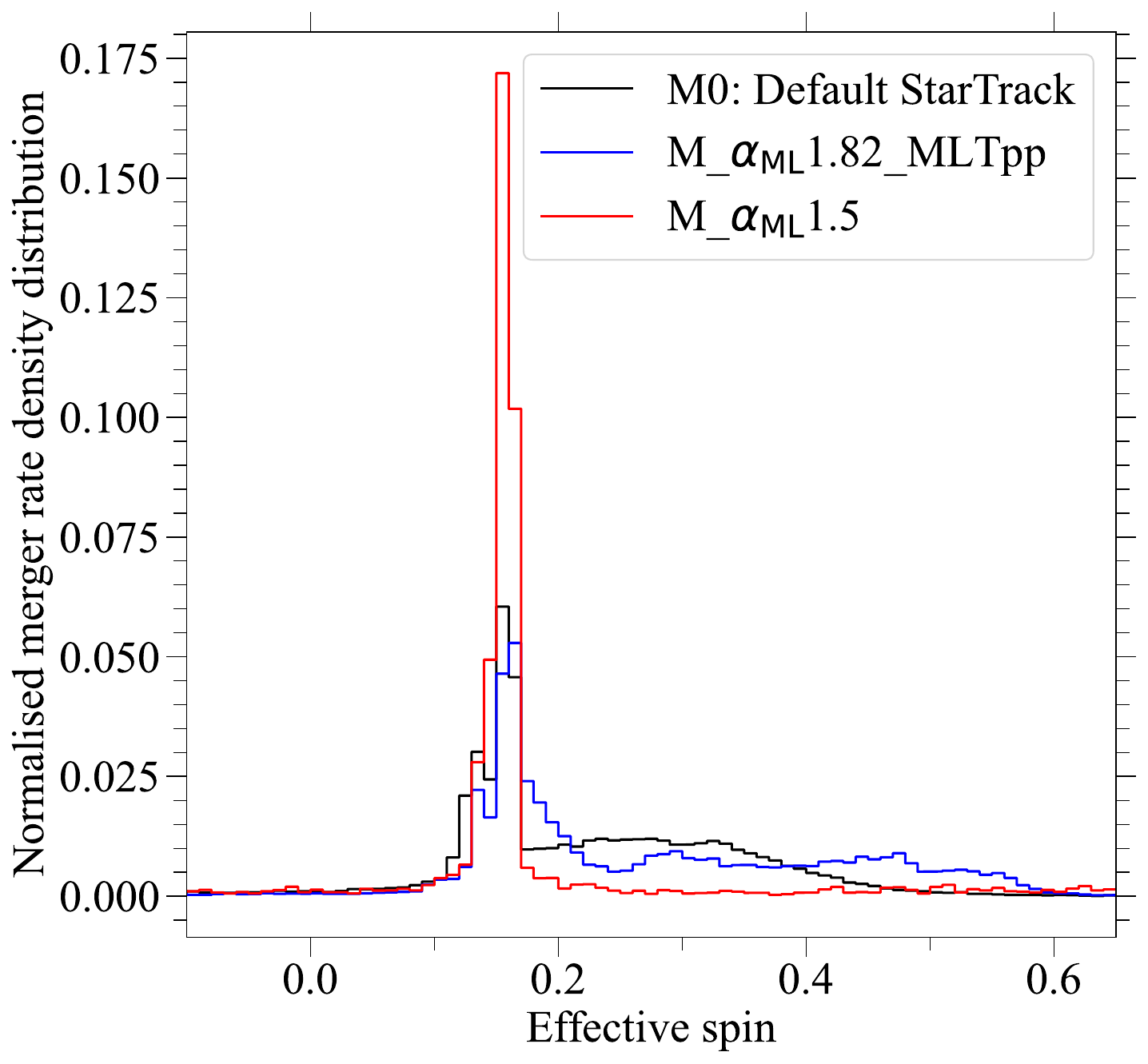}
    \caption{Normalised spin distribution of the BH-BH mergers within redshift 2. The reference M0 model for {\tt StarTrack} and M\_$\alpha_{\rm ML}$1.82\_MLTpp show a spread to positive values, while M\_$\alpha_{\rm ML}$1.5 concentrates uniquely on the peak at 0.15.
    }
    \label{fig:BHBH_spin}
\end{figure}

\paragraph{\underline{Spin Distribution}}
Figure \ref{fig:BHBH_spin} shows the spin distribution for the BH-BH merger according to our models. While both M0 and M\_$\alpha_{\rm ML}$1.82\_MLTpp models exhibit a distribution extending to positive values due to tidal interactions between WR stars and BHs \citep{Belczynski_2020,Olejak_2020b,Bavera_2022}, M\_$\alpha_{\rm ML}$1.5 lacks a significant population of high-spin mergers. This is due to the reduced formation rate of WR-BH binaries with sufficiently close orbits capable of producing substantial tidal spin-up. In contrast, most BH-BH mergers with effective spins below 0.15 originate from binaries retaining primordial angular momentum rather than acquiring spin from WR-BH interactions \citep{Olejak_2020b,Olejak_2024}. The uncertainties surrounding CE parameters ($\alpha_{\rm CE}$ and $\lambda_{\rm CE}$) hinder precise predictions of post-CE orbital separations, making it challenging to definitively distinguish between WR-BH binaries that undergo significant spin-up and those that do not. Alternative channels such as the SMT scenario can contribute to the observed effective spin distribution \citep{Pavlovskii_2017,Gallegos_2021,Olejak_2021,Broekgaarden_2022,Dorozsmai_2024,Olejak_2024}.

\subsection{Relation between BH-BH mass and progenitor metallicity}
\label{subsec:BHBH_MRD_M_Z}

\begin{figure}
  \centering
    \includegraphics[width=1\columnwidth,height=7.4 cm]{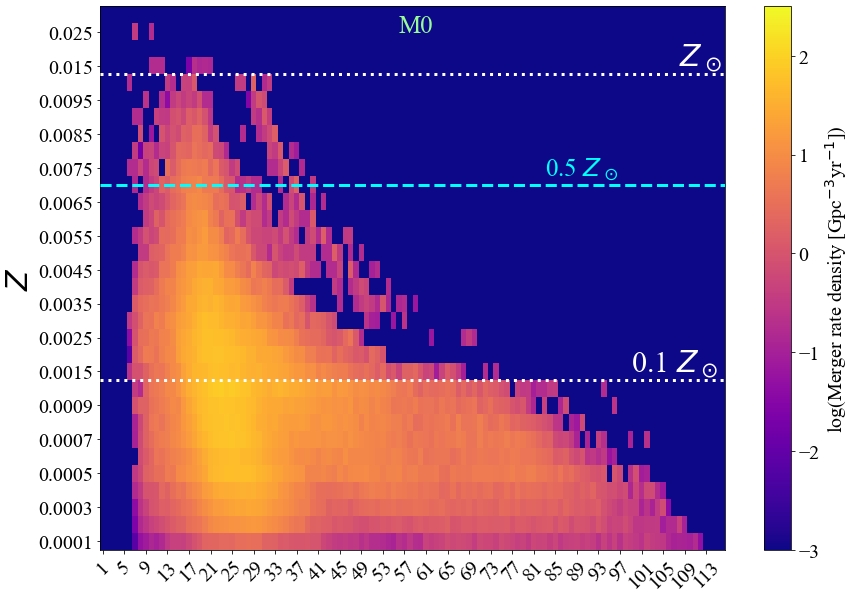}
  \quad
    \includegraphics[width=1\columnwidth,height=7.4 cm]{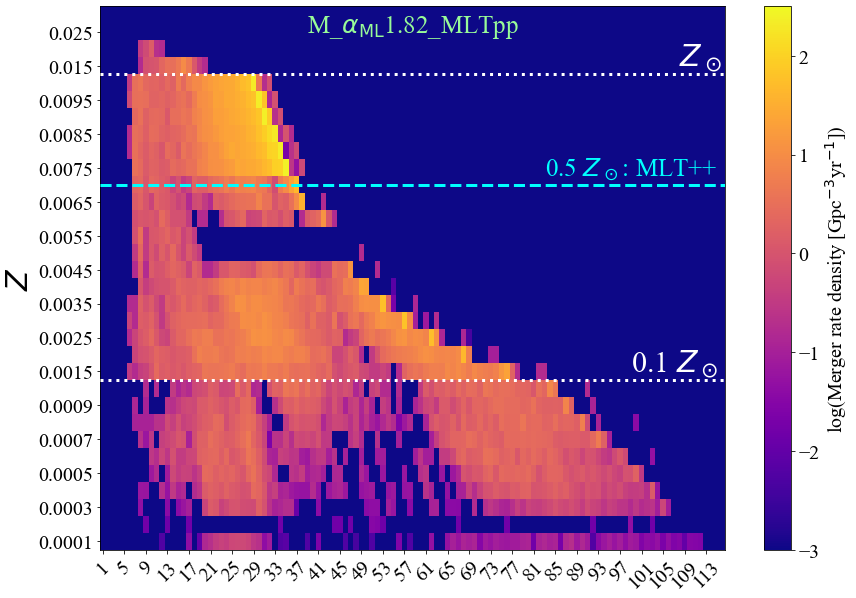}

    \includegraphics[width=1\columnwidth,height=7.4 cm]{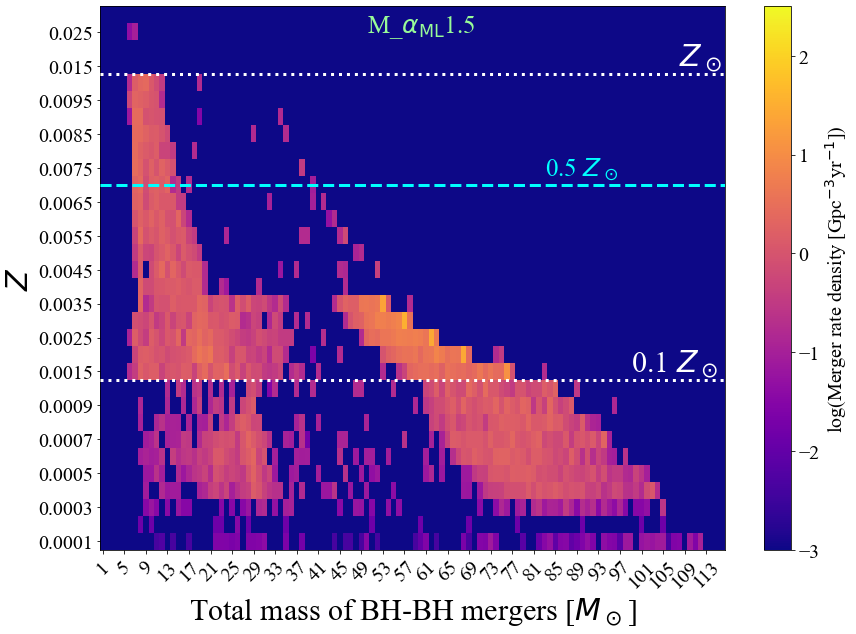}
  
  \caption{BH-BH total mass distribution within redshift 2 as a function of metallicity for M0 (top), M\_$\alpha_{\rm ML}$1.82\_MLTpp (middle), M\_$\alpha_{\rm ML}$1.5 (bottom). The dashed cyan line represents the metallicity threshold beyond which the formation of a convective envelope is estimated from the {\tt MESA} tracks using \textit{MLT++} for M\_$\alpha_{\rm ML}$1.82\_MLTpp (more details in Section~\ref{subsec:uncert}).}
  \label{fig:rMZ}
\end{figure}

In Figure~\ref{fig:rMZ} we show within redshift 2 ($z$\,<\,2) the relation between BH-BH merger rate density, total mass, and initial metallicity of the progenitor binaries. 

\paragraph{\underline{M0}} Most of the contribution comes from the low-mass end of the BH-BH mergers and from metallicities $Z\lesssim$\,0.5\,$Z_\odot$. The BH-BH total mass peaks at $\sim$21~$M_\odot$ with no significant metallicity-dependent shift.

\paragraph{\underline{M\_$\alpha_{\rm ML}$1.82\_MLTpp}} In contrast to M0, the peak of the mass distribution shifts towards more massive mergers (see also Figure~\ref{fig:BHBH_mass}). This trend exhibits a metallicity dependence. The key here is the inclusion of binaries containing HG donors with convective envelopes in our model. Figure~\ref{fig:HR_conv} shows that only the most massive or luminous (i.e. with a more massive core) donors can develop a convective envelope during their HG phase, which explains the contribution to more massive BH-BH mergers. The \textit{MLT++} method broadens the parameter space for CE ejection, significantly boosting the contribution to the BH-BH merger rate for $Z$\,$\geq$0.5~$Z_\odot$.

\paragraph{\underline{M\_$\alpha_{\rm ML}$1.5}} Compared to M\_$\alpha_{\rm ML}$1.82\_MLTpp we show a decrease of BH-BH mergers at $Z$\,<\,0.5~$Z_\odot$. This difference stems primarily from the varying $\alpha_{\rm ML}$ values between the models. However, a more significant disparity emerges in the BH-BH merger rates at $Z\geq$\,0.5~$Z_\odot$. Here, M\_$\alpha_{\rm ML}$1.5 shows a drastic decline in BH-BH mergers of mass between $\sim$21~$M_\odot$ and 50~$M_\odot$. While the $\alpha_{\rm ML}$ difference still contributes, the primary driver for this sharp drop is the choice not to suppress density inversion effects with \textit{MLT++} in super-Eddington regions of stellar envelopes at these metallicities. This distinction is reflected in the considerably different $T_{\rm eff}$ thresholds seen in the right panel of Figure~\ref{fig:HR_conv}.

\paragraph{\underline{Almost no BH-BH mergers at $Z\gtrsim$\,$Z_\odot$}} In all models we find a nearly-total absence of BH-BH mergers at $Z\gtrsim$\,$Z_\odot$, which was also reported in other population studies (e.g.\citealt{Belczynski_2010,Cason_2024}). We consider that a contributing factor for this phenomenon might be metallicity-dependent wind-driven mass loss, which narrows the $M_{\rm ZAMS}$ range at which a collapse into a BH happens to only the most massive (and therefore rarest) stars. 

\subsection{Formation of T\.ZOs and QSs}

For each of our models (Section \ref{subsubsec:formations}) we extrapolate our synthetic population of T\.ZOs/QSs to encompass the whole latest star formation event in the Galactic disk. We estimate the numbers of T\.ZOs and QSs and their formation rates (Table \ref{tab:HS}). We show the almost complete absence of T\.ZO/QS in the Galactic disk for the Strong RSG Pulsation scenario. For the Weak RSG Pulsation scenario the T\.ZO and QS population increases for all the models, but never goes beyond respectively 23 and 8 objects. At $Z_\odot$ we predict a formation rate of 2-3\,$\times$\,10$^{-4}$~yr$^{\rm -1}$ for T\.ZOs, which is comparable to the results from the literature \citep{Cannon_1993,Podsiadlowski_1995,Hutilukejiang_2018,Nathaniel_2024}. This suggests that even under optimistic assumptions their low formation rate, in addition to their short lifetimes, makes their detectability unlikely. 
On top of that, \cite{Farmer_2023} predicts that T\.ZOs can be observationally distinguishable from RSGs only at low metallicities ($Z$\,<\,$10^{\rm -3}$), which are rare in the considered star-forming population of the Galactic disk.

\paragraph{\underline{Are the lifetimes for QSs overestimated?}}
We stress that assuming that the timescale relation for T\.ZOs is also valid for QSs is a speculative approach. The accretion-induced radiation pressure may lead to completely block nuclear burning inside a QS due to the cooling of stellar interiors \citep{Bellinger_2023}. Once the photospheric temperature drops to a specific threshold, which is a function of metallicity, the opacity drastically decreases and in turn the temperature reaches a condition at which no hydrostatic solution can be found for the QS convective envelope. After this temperature is reached, QSs may quickly disperse by radiation pressure \citep{Begelman_2008}, which would further narrow the temporal window at which such objects are detectable. On the other hand, \citep{Bellinger_2023} showed that at high $M_{\rm star}$/$M_{\rm BH}$ ratios the lifetime of a QS is almost unchanged if compared with a non-QS MS star of the same mass. We nevertheless argue that this case would only be likely for Hawking stars \citep{Hawking_1971,Bellinger_2023}, which are QSs with a primordial BH in their centre that can be of sub-Solar mass. In the case of QSs formed from massive binaries during a CE event the lifetimes cannot be comparatively as high because i) the central BHs are at least 2.5~$M_\odot$ ii) most CE events with a BH accretor in our population are initiated with a post-MS donor star. Considering the number of QSs (1 at most; Table~\ref{tab:HS}) we estimate to reside in the Galaxy, it is unlikely that such objects can be detected. However, QSs outside our Galaxy may be potentially detected through X-ray observations in case of a thin-envelope QS, where a BH efficiently accretes mass through a disk \citep{HutchinsonSmith_2023}. This disk provides accretion support to the optically-thin envelope and such an object can be observed as an ultraluminous X-ray source \citep{Everson_2024}. If thin-envelope QSs were to exist, at the moment there is nevertheless no understanding on how we could distinguish them from other ultraluminous X-ray sources.

\begin{table*}[t!]
\centering
\caption{
Estimated population of Thorne-\.Zytkow objects (T\.ZOs) and quasi-stars (QSs) in the Galactic thin disk at the present time, and their respective formation rate at $Z_\odot$. 
}
\label{tab:HS}
\begin{tabular}{l|cc|cc|cc}
\hline
\multirow{2}{*}{Model} & \multicolumn{2}{c|}{Strong RSG Pulsation} & \multicolumn{2}{c|}{Weak RSG Pulsation} & \multicolumn{2}{c}{Formation rate [10$^{-5}$ yr$^{-1}$]} \\
 & T\.ZO & QS & T\.ZO & QS & T\.ZO & QS \\
\hline
M0 & 0 & 1 & 22 & 8 & 29.5 & 3.5 \\
M0$_{\rm RMAX}$ & 0 & 0 & 19 & 3 & 23.0 & 2.4 \\
\hdashline
M\_$\alpha_{\rm ML}$1.5 & 1 & 0 & 21 & 2 & 29.5 & 2.6\\
M\_$\alpha_{\rm ML}$1.5$_{\rm RMAX}$ & 1 & 0 & 20 & 3 & 29.5 & 2.6 \\
\hdashline
M\_$\alpha_{\rm ML}$1.82\_MLTpp & 0 & 0 & 21 & 1 & 29.4 & 2.6 \\
M\_$\alpha_{\rm ML}$1.82\_MLTpp$_{\rm RMAX}$ & 0 & 1 & 21 & 0 & 29.4 & 2.5 \\
\hline
\end{tabular}
\end{table*}

\subsection{Uncertainties in 1D modeling of stellar envelopes}
\label{subsec:uncert}

\paragraph{\underline{Stellar expansion}}
1D stellar models have inherent uncertainties.  For instance \cite{Grafener_2012,Agrawal_2022,Agrawal_2022b,Romagnolo_2023} showed how different assumptions about stellar physics and the use of different evolutionary codes can lead to significantly diverging results for stellar envelope expansion. In particular, the evolution of stars at $M_{\rm ZAMS}$\,$\gtrsim$\,50~$M_\odot$ remains poorly constrained, with radial expansion that can change by orders of magnitude \citep{Romagnolo_2023}. This effect is considerable in the fitting formulae used in our population synthesis models \citep{Hurley_2000}, since we extrapolate them for any star above 50~$M_\odot$ and interpolate them for any metallicity different from the ones used in \citep{Pols_1998} tracks ($Z$\,=\,0.0001, 0.0003, 0.001, 0.004, 0.01, 0.02, and 0.03). For even more massive stars, e.g. \cite{Maeder1989,Ekstrom_2012,Yusof_2022,Bavera_2023,Romagnolo2024,Vink_2024} have shown that stars at  $M_{\rm ZAMS}$\,$\gtrsim$\,100~$M_\odot$ and $Z$\,$\gtrsim$\,$Z_\odot$ only display a moderate or negligible expansion due to the combined effect of strong winds and efficient internal mixing. Choosing these models as the baseline for our estimates would likely have considerably narrowed the window for massive binaries to start or evolve past CE events and become GW sources. A major source of uncertainty for stellar radius evolution is the Humphreys-Davidson (HD) limit \citep{Humphreys_1994}. This limit marks a zone in the HR diagram where stars are rarely, if ever, observed \citep{Humphreys_1979,Humphreys_1994,Ulmer_1998,Davies2018}. This boundary seems to be a HR region where many RSGs with convective envelopes are predicted to be (see Figures~\ref{fig:HR_conv} and \ref{fig:RMAX_HR}, and e.g. \citealt{Gilkis_2021}). This may indicate that stars do not expand enough to enter that HR region, which can be an important calibration point for stellar models.

\paragraph{\underline{Convective envelope development}}
In terms of the formation of convective envelopes model uncertainties are large as well. The major difference between M\_$\alpha_{\rm ML}$1.5 and M\_$\alpha_{\rm ML}$1.82\_MLTpp, at least at high metallicities, depends on the treatment of convection in superadiabatic envelopes nearing the Eddington limit. This phenomenon is associated with the presence of sub-surface opacity peaks arising from the ionization of iron or helium (depending on the surface temperature, \citealt{Cantiello_2009,Jermyn_2022}), combined with low density. When the peak is located deep within the star, convection operates efficiently. This efficient convection can redistribute material within the star, leading to a nearly constant entropy profile and eliminating the density inversion. However, if the iron opacity peak is situated in an outer low-density envelope region, the convective flux becomes less effective. The 1D stellar structure adjusts accordingly, leading to (possibly artificial) density inversions. These subsurface effects affect the global properties of the star (e.g. radius and temperature) and in turn the $T_{\rm eff}$ threshold for convective envelope development explored in this paper. 3D simulations by \cite{Jiang_2015} suggest that "stellar engineering" solutions (\textit{MLT++} or \textit{use\_superad\_reduction}; \citealt{Jermyn_2023}) to suppress density inversion might only be motivated for sub-sonic turbulence in convective layers. This is because shocks can form in these regions, leading to density fluctuations that methods such as mixing length theory \citep{BohmVitense_1958} cannot handle \citep{Joyce_2023}. This means that in 1D models it is still unclear how and when density inversion should be artificially suppressed. Regardless of whether (and when) density inversion is a realistic phenomenon within near-Eddington regions of stellar envelopes or just an artifact of 1D stellar models, one of the major implications arising from its suppression is a steeper temperature gradient, which leads to the formation of a convective envelope at smaller radii. This can be seen in Figure~\ref{fig:HR_conv} for the $Z_\odot$ plot, where the blue line from M\_$\alpha_{\rm ML}$1.82\_MLTpp, which represents the {\tt MESA} model with a density suppression prescription, shows the development of a convective envelope at considerably higher $T_{\rm eff}$ than what M\_$\alpha_{\rm ML}$1.5, that allows instead for density inversion within stellar envelopes, prescribes (red line).

\paragraph{\underline{Binding Energy}}

In this study, we employ the presence of a convective envelope in massive CE donors as a proxy for a steep density profile at the core-envelope boundary and low envelope $E_{\rm bind}$ prior to any CE outcome estimate through the $\alpha\lambda$ formalism (the $\lambda$ values are still calculated with the \citealt{Dominik_2012} model), assuming that all internal and recombination energy contributes to the envelope unbinding. We emphasize, however, that for stars developing convective envelopes during the HG phase (i.e. when CHeB has not yet started) $E_{\rm bind}$ might remain sufficiently high to prevent CE ejection \citep{Klencki_2021}. Moreover, our convective envelope models stem from single-star {\tt MESA} simulations, introducing substantial uncertainties given the likely past impact of companions on CE donors. For example, \cite{Nathaniel_2024} found that nearly 92\% of their synthetic CE donors were rejuvenated from a past mass accretion phase. While the influence of accretion on envelope unbinding remains unclear, it is likely to affect the $E_{\rm bind}$ \citep{Renzo_2023,Landri_2024b}. Similarly, the potential SN when the accretor collapses into a CO can modify $E_{\rm bind}$. Beyond mass stripping, SNe in short-period binaries may inject energy deep into the companion’s envelope, inducing inflation and therefore changes in stellar structure \citep{Ogata_2021,Hirai_2023}.

\paragraph{\underline{Luminous Blue Variables}}
Another factor that we didn't consider in our {\tt MESA} models is that the hydrodynamical turbulence and shocks within superadiabatic layers might be responsible for extreme mass loss events in luminous supergiants and the LBV phenomenon \citep{Owocki_2004,Vink_2011,vanMarle_2008,Mennekens_2014,Bestenlehner_2014,Quataert_2016,Grafener_2021,Cheng_2024,Mukhija_2024}. The existence of the HD limit might be at least partially due to stars nearing their Eddington limit \citep{Humphreys_1979,Ulmer_1998,Romagnolo2024}.

\section{Conclusions}

Various population synthesis models of gravitational wave sources from common-envelope evolution are based on assumptions that are in tension with recent results underlying the crucial role of convective envelopes for successful common envelope evolution \citep{Kruckow_2016,Klencki_2020,Klencki_2021,Marchant_2021} In our study, we employed 1D stellar evolution models ({\tt MESA}) to investigate the conditions under which massive giants develop convective envelopes, considering varying assumptions regarding mixing length and the treatment of density inversions in super-Eddington regions. We derived analytical formulae, based on our {\tt MESA} simulations, that define an effective temperature threshold for convective envelope formation as a function of metallicity and luminosity. These formulae were then integrated into the {\tt StarTrack} rapid population synthesis code and used to update the common envelope treatment to require the presence of an outer convective envelope for a successful common envelope ejection. This revised treatment improves upon the previous approach (e.g. \citealt{Belczynski_2020} with {\tt StarTrack}, \citealt{vanSon_2022} with {\tt COMPAS}) in which such a pre-requisite for common envelope ejection was based on the evolutionary type of the giant donor (namely, Hertzprung-Gap donors were often excluded, following \citealt{Hurley_2000} nomenclature). Finally, we performed a population synthesis study of gravitational wave sources from the common envelope channel.  Under the assumption that any failed common envelope ejection with a neutron star or black hole accretor leads to the formation of respectively a Thorne-$\dot {\rm Z}$ytkow Object or a quasi-star, we expanded our study to estimate their Galactic population and formation rates.

\paragraph{\underline{Most common envelope events are with radiative envelopes}} 
Massive stars tend to develop convective envelopes only during advanced stages of their evolution and radial expansion (Figure~\ref{fig:HR_conv}). Through our population model, we find that the majority of common envelope events are initiated earlier, by giant donors with radiative envelopes. In prior models, these events would often lead to a common envelope ejection and the formation of a gravitational-wave source. Here, with our revised model, we include a pre-requisite of an outer convective envelope for common envelope ejection. This leads to a notable decrease (up to a factor of 20, Table~\ref{tab:MRD}) in the yield of the common envelope channel for the BH-BH merger rates mergers with a total mass between $\sim$20~$M_\odot$ and 50~$M_\odot$. Unexpectedly, we find the merger rate density to increase for BH-BH mergers with total mass $M_{\rm tot}$\,$\gtrsim$\,50~$M_\odot$ (depending on the model). This is because many of these sources form through common envelope events with donors that are classified as stars of the Hertzsprung-gap type (following \citealt{Hurley_2000}) and that were deemed to lead to common envelope mergers. Here, we show that many of such donors may in fact be convective-envelope giants and good candidates for the CE channel. 

\paragraph{\underline{BH-BH total mass distribution}}

In our new models there is a drop in the merger rates starting from BH-BH mergers up to $\sim$50~$M_\odot$ (Figures~\ref{fig:BHBH_mass} and \ref{fig:rMZ}). Above that mass, we find that the rate is comparable or higher than the prior models due to Hertzsprung-gap donors developing convective envelopes and therefore contributing to the population of gravitational wave sources. The net result of this is that the total mass distribution changes from a declining power law to a bimodal distribution. We note, however, that our criteria for what we treat as an convective-envelope star is rather optimistic, setting a threshold at just 10\% of the envelope mass being in outer convective regions. With the increasing stellar mass, the subsurface convective zones grow bigger, such that the 10\% threshold can be met further away from the classical Hayashi line of convective stars. A more conservative threshold of 30-50\% would likely lead to a considerable decrease of the CE channel yield for BH-BH mergers with $\sim$50~$M_\odot$ and beyond.

\paragraph{\underline{Thorne-\.Zytkow objects and quasi-stars}}
At $Z$\,$\approx$\,$Z_\odot$, under the optimistic assumption that mergers during common envelope events with black hole or neutron star accretors lead to the formation of quasi-stars and Thorne-$\dot{\rm Z}$ytkow objects rather than transient events, their estimated formation rate (respectively $\sim$\,2.6\,$\times$\,10$^{-5}$ and 2.9$\,\times$\,10$^{-4}$~yr$^{\rm -1}$; Thorne-$\dot{\rm Z}$ytkow Object rates in agreement with \citealt{Cannon_1993,Podsiadlowski_1995,Hutilukejiang_2018,Nathaniel_2024}) remains almost unaltered throughout all the considered models. However, even with a clear method for distinguishing such objects from giant stars, their limited lifetime ($\sim$10$^5$~yr) makes them unlikely to be detected in the Milky Way.

\paragraph{\underline{Key uncertainties}}

Predicting when massive giants develop convective envelopes is very sensitive to the treatment of radiation-dominated outer envelope layers that approach the Eddington limit. Ad-hoc simplified solutions already exist in 1D codes, such as \textit{MLT++} \citep{Paxton2013}, \textit{use\_superad\_reduction} \citep{Jermyn_2023} or super-Eddington mass ejection \citep{Cheng_2024,Mukhija_2024}. This results in considerable differences in the efficiency of the common envelope channel, to the point that models suppressing superadiabaticity can result in up to one order of magnitude more BH-BH mergers than the ones that do not. Ultimately a more comprehensive guidance on how to improve 1D codes will hopefully come from 3D simulations \cite[e.g.,][]{Jiang_2015}. In binary population models, most common envelope events leading to gravitational-wave sources are initiated by stars that are above the Humphrey-Davidson limit. Understanding the origin of this limit and in particular the nature of stellar winds and eruptive mass loss from the most luminous supergiants is one of the key challenges in refining the isolated binary evolution scenario. In particular, it has been shown that strong mass loss can prevent very massive stars from expanding \citep[e.g.,][]{Maeder1989,Ekstrom_2012,Bavera_2023,Cheng_2024,Mukhija_2024,Romagnolo2024,Vink_2024} and therefore contributing to the common envelope channel as stellar donors.

\paragraph{\underline{Are the BH-BH merger rates from the common envelope} \\\underline{channel overestimated?}}
While in our new models the common envelope channel can reproduce the observed rate of NS-NS mergers and likely contributes to the population of low-mass BH-BH mergers, we find that the yield of the common envelope channel may have been so far been overestimated for the formation of massive BH-BH mergers ($M_{\rm tot}$\,$\gtrsim$\,20~$M_\odot$). Our results suggest that BH-BH merger rate density can be as low as 3~Gpc$^{\rm -3}$\,yr$^{\rm -1}$, which is roughly 22 times lower than what is estimated with the standard models ($\sim$66~Gpc$^{\rm -3}$\,yr$^{\rm -1}$). Notably, in the new models the rate of the most massive BH-BH mergers ($M_{\rm tot}$\,$\gtrsim$\,50~$M_\odot$) remains high (Figures~\ref{fig:BHBH_mass} and \ref{fig:rMZ}) due to the inclusion of Hertzsprung-gap donors that were previously disregarded for common envelope evolution (see Section~\ref{subsubsec:CE} for more detail). BHs this massive are expected to originate from stars with masses $M$\,$\gtrsim$\,40~$M_\odot$ that would lie above the Humphreys Davidson limit. It is unclear whether such stars ever expand sufficiently to develop outer convective envelopes and why no such red supergiants have been observed. With a more detailed implementation of stellar winds that can account for the initiation of strong optically thick winds (e.g. \citealt{Chen_2015,Sabhahit_2022,Sabhahit_2023,Romagnolo2024}; transition criteria for thick winds from e.g.\citealt{Vink_2012} and \citealt{Bestenlehner_2020}), the rate of massive BH-BH mergers ($M_{\rm tot}$\,$\gtrsim$\,50~$M_\odot$) from common envelope evolution is likely to significantly decrease.

\begin{acknowledgements}
AR, JK, and AVG would like to dedicate this article to Krzysztof Belczyński who sadly passed away when the work was still in progress. We thank him for years of collaboration, honest mentorship, and the inspiring energy and passion that will never be forgotten. We thank T. Braun, A. C. Gormaz-Matamala, M. Renzo, R. Hirai, S. Justham, A. Olejak, I. Kotko, K. Nathaniel and T. Shenar for the useful discussions. We thank the Kavli Foundation and the Max Planck Institute for Astrophysics for supporting the 2023 Kavli Summer Program during which much of this work was completed. AR acknowledges the support from the Polish  National  Science  Center  (NCN)  grant  Maestro  (2018/30/A/ST9/00050). AR and JK acknowledge the support of a ESO SSDF grant.  JK acknowledges the support of a ESO fellowship. Special thanks go to the tens of thousands of citizen-science project ``Universe@home" (universeathome.pl) enthusiasts that help to develop the {\tt StarTrack} population synthesis code used in this study. AVG acknowledges funding from the Netherlands Organisation for Scientific Research (NWO), as part of the Vidi research program BinWaves (project number 639.042.728, PI: de Mink).
\end{acknowledgements}

\bibliographystyle{aa} 
\bibliography{aanda.bib} 

\appendix

\section{Outer convective envelope evolution}
\label{sec:outconvenv}

\begin{figure}[h]
    \centering
    \includegraphics[width=0.495\textwidth]{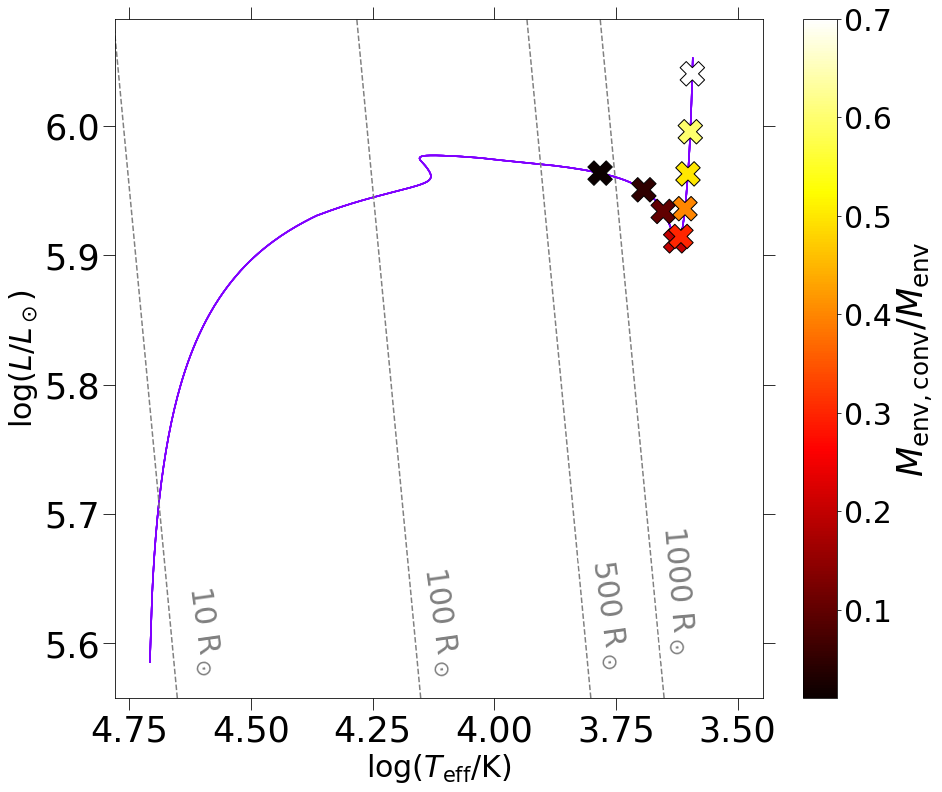}
    \includegraphics[width=0.495\textwidth]{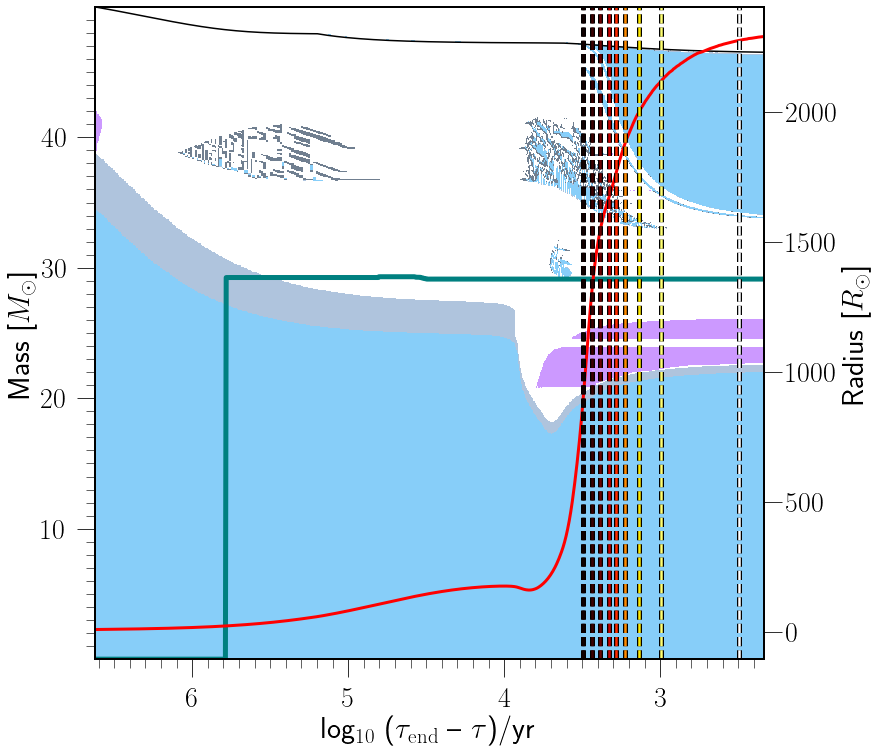}
    \caption{Evolution of a $M_{\rm ZAMS}$\,=\,50~$M_\odot$ star at 0.1~$Z_\odot$ from M\_$\alpha_{\rm ML}$1.82\_MLTpp until $\sim$~70\% of its envelope mass becomes convective. Top panel: HR diagram. We show the HR position of the star when $M_{\rm env, conv}$ is 1, 5, 10, 20, 30, 40, 50, 60 and 70\% of $M_{\rm env}$. Bottom panel: Kippenhahn diagram, with radial evolution represented by the red line. In cyan, purple and gray respectively the convective, semiconvection and overshooting regions, while the green line represent the post-MS core-envelope boundary. The dashed vertical lines have the same colors of the scatter points in the left panel and represent the evolutionary stage at which the star reaches its respective $M_{\rm env, conv}$/$M_{\rm env}$ ratio.}
    \label{fig:kipp_50Msun}
\end{figure}

Figure~\ref{fig:kipp_50Msun} shows a Kippenhahn and an HR diagram of a $M_{\rm ZAMS}$\,=\,50~$M_\odot$ star at 0.1~$Z_\odot$ taken from M\_$\alpha_{\rm ML}$1.82\_MLTpp. To exemplify the evolution of its convective envelope we evolved the model until its outer convective envelope mass $M_{\rm env, conv}$ reaches $\sim$70\% of the total envelope mass $M_{\rm env}$. If such a star was a CE donor, we prescribe that it would have a low enough $E_{\rm bind}$ for its host binary to eject the CE only once it will expand above $\sim$\,1700~$R_\odot$. 

\begin{figure}[h]
    \centering
    \includegraphics[width=0.5\textwidth]{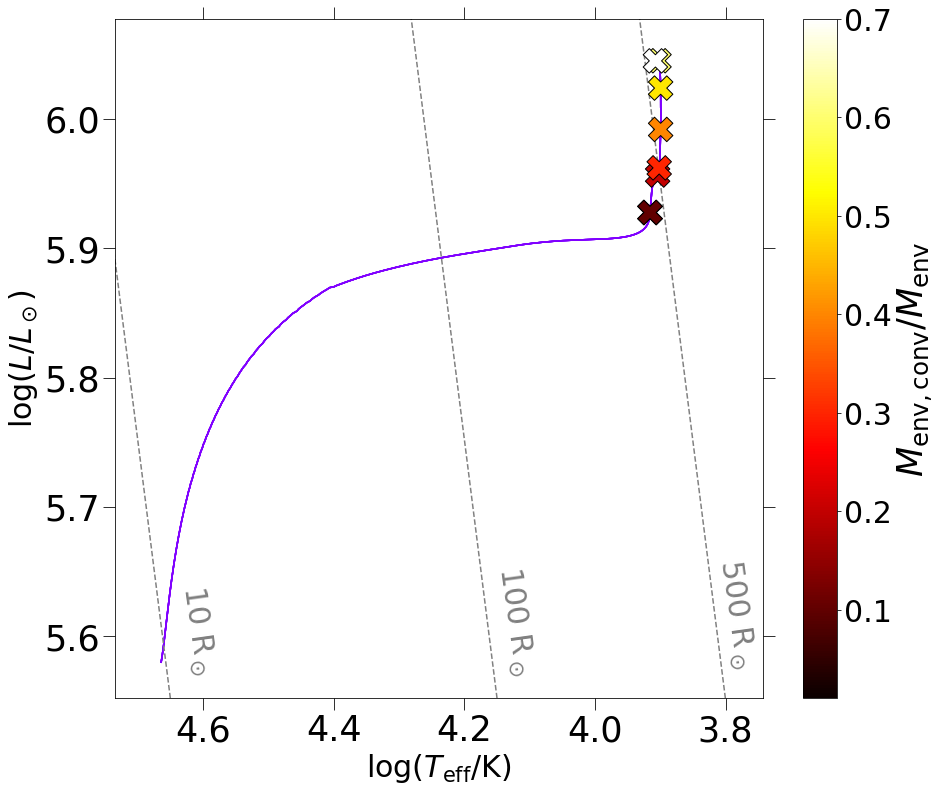}
    \includegraphics[width=0.5\textwidth]{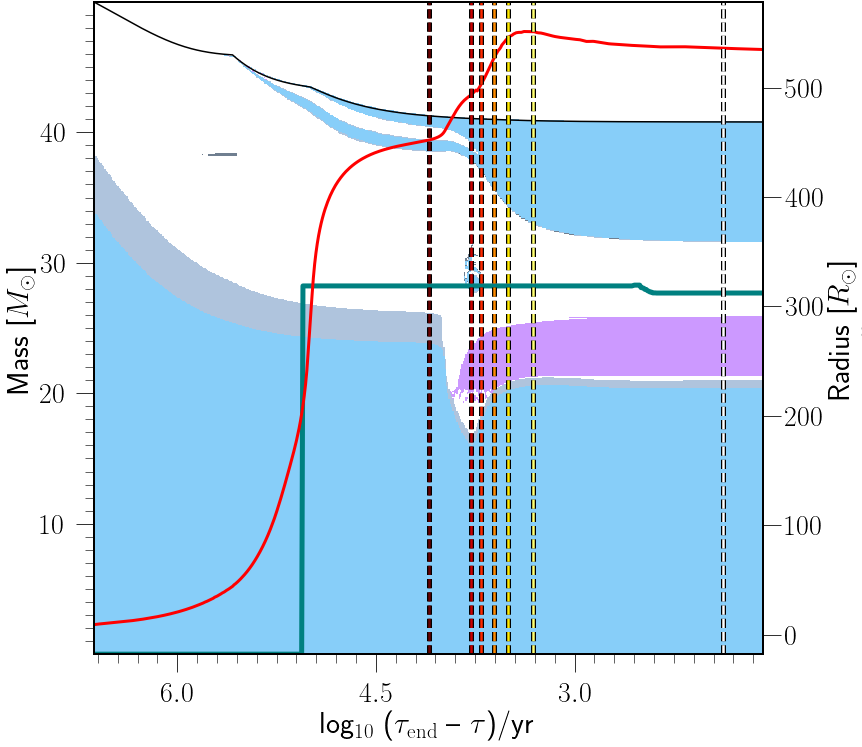}
    \caption{Same as Figure~\ref{fig:kipp_50Msun}, but at $Z$\,=\,$Z_\odot$. and with the HR position of the star when $M_{\rm env, conv}$ is 10, 20, 30, 40, 50, 60 and 70\% of $M_{\rm env}$.}
    \label{fig:kipp_50MsunZsun}
\end{figure}

Figure~\ref{fig:kipp_50MsunZsun} shows the same plots of Figure~\ref{fig:kipp_50Msun}, but for a star at $M_{\rm ZAMS}$\,=\,50~$M_\odot$ star at $Z_\odot$. We highlight that for this metallicity the \textit{MLT++} procedure is enabled, which can be seen from the higher $T_{\rm eff}$ at which a convective envelope forms. The formation of a convective envelope is quicker than the star at 0.1~$Z_\odot$. Once such an envelope forms for the first time, it already possesses 10\% of the total envelope mass. For such a star to form a convective envelope of at least 10\% of the total stellar mass, it would need to expand at least at $\sim$400~$R_\odot$.

\section{Maximum stellar expansion}
\label{sec:RMAX}

\subsection{Metallicity dependence}

\begin{figure}[h]
\includegraphics[width=0.465\textwidth]{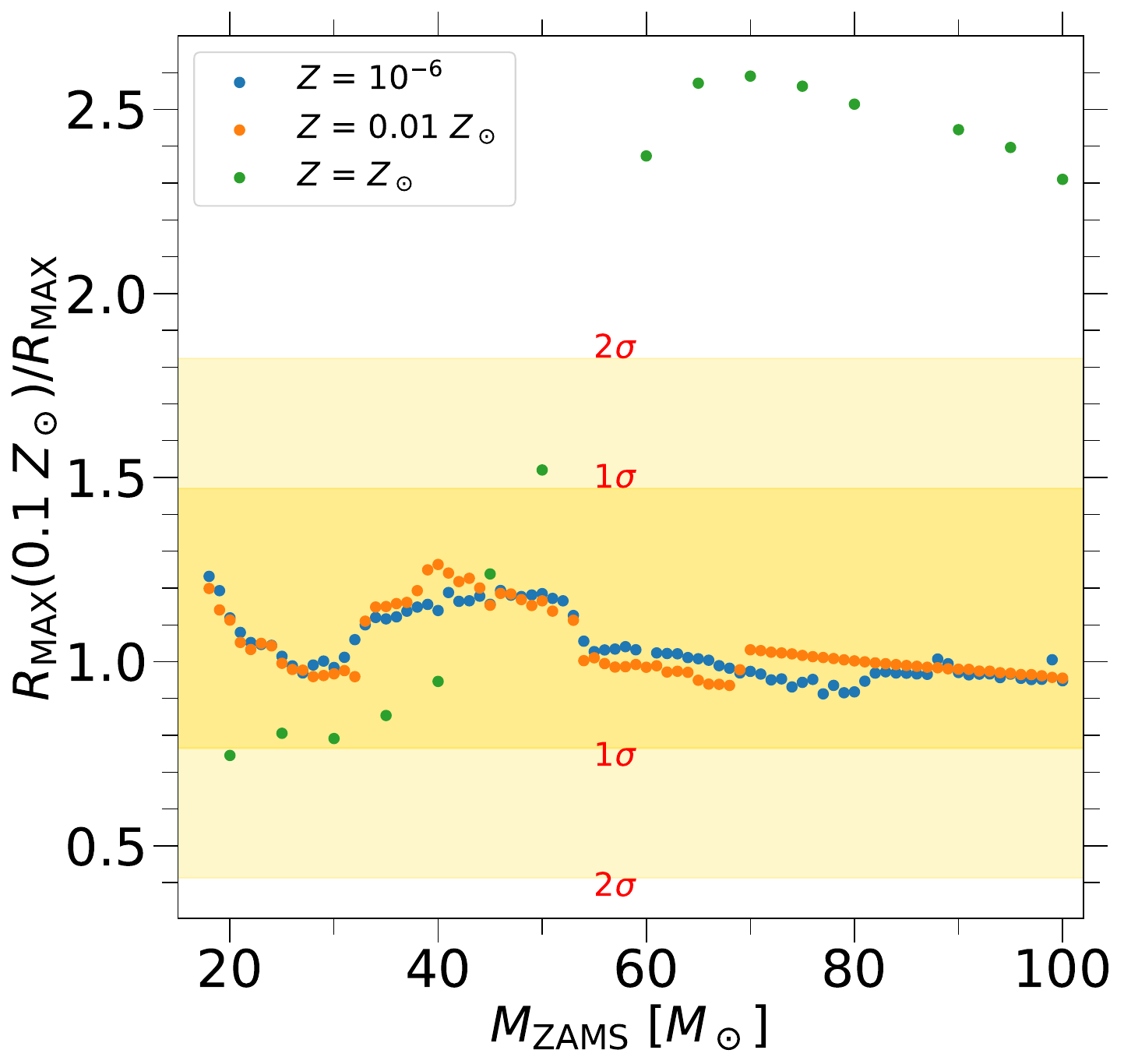}
\caption{
Maximum radius $R_{\rm MAX}$ ratios as a function of ZAMS mass $M_{\rm ZAMS}$ between the reference metallicity $Z$ = 0.1~$Z_\odot$ and the other adopted metallicities according to our {\tt MESA} simulations. We show the $R_{\rm MAX}$(0.1~$Z_\odot$/$R_{\rm MAX}$ mean value at $\sim$1.1 and its Gaussian distribution up to 2$\sigma$. The variability of all the low-Z models with respect to  0.1~$Z_\odot$ sits within 1$\sigma$, while it gets significantly beyond 2$\sigma$ at  $Z_\odot$ for $M_{\rm ZAMS} \gtrsim$ 60 $M_\odot$ due to the increasingly large effect of wind mass loss and the treatment of superadiabaticity.
}
\label{fig:RMAX_Z}
\end{figure}

Figure \ref{fig:RMAX_Z} illustrates the impact of varying metallicities on stellar expansion in our {\tt MESA} simulations. We analyze the change in the $R_{\rm MAX}$ parameter as a function of initial stellar mass, comparing it between the reference model at $Z$\,=\,0.1~$Z_\odot$ from \cite{Romagnolo_2023} and models with lower and higher metallicities (Z = $10^{-6}$, 0.01~$Z_\odot$, $Z_\odot$). For $Z$\,=\,$Z_\odot$ we adopt the same adjustments for superadiabaticity and wind-driven mass loss that were described in Section~\ref{subsubsec:conv_env}, i.e. the use of the module \textit{MLT++} and \textit{Dutch\_scaling\_factor}\,=\,0.5. We explore the difference in metallicities as the ratio between $R_{\rm MAX}$($M_{\rm ZAMS})$ at 0.1~$Z_\odot$ and the $R_{\rm MAX}$($M_{\rm ZAMS})$ from the other metallicities.

We observe small discrepancies among the models at $Z$\,$\leq$\,0.1~$Z_\odot$, as all the variations fall within a 1 sigma $\sigma$ deviation from the mean value. Furthermore, the mean value itself remains in close proximity to unity, approximately 1.1, indicating the relatively modest impact of metallicity on $R_{\tt MAX}$ on the considered metallicities. However, a substantial difference arises from the $Z_\odot$ model. This difference is moderate for stars with mass $\lesssim$\,50~$M_\odot$. Above this mass, the disparities become considerable. This is due to the extreme mass loss that is characteristic to massive stars at high metallicity. It is worth emphasizing that even in the context of low-metallicity scenarios, massive stars can already expand up to 2.5 times the $Z_\odot$ counterparts despite our deliberate reduction of wind mass loss by half for the high-metallicity scenarios. This is due to the fact that at this metallicity massive stars still lose more envelope mass due to winds and because the use of \textit{MLT++} reduces their expansion (more details in Section~\ref{subsec:uncert}). If we kept the wind mass loss to its default values, the differences would be even more striking, since even more envelope mass would have been lost.

For $Z_\odot$ metallicities, we therefore fitted the maximum expansion of stars from our {\tt MESA} simulations to the formula below:

\begin{align}
    \log (R_{\rm MAX}) &= -3.46\times 10^{-9}M_{\rm ZAMS}^5 + 8.92\times 10^{-7}M_{\rm ZAMS}^4 \nonumber \\
    &\quad -7.76\times10^{-5}M_{\rm ZAMS}^3 + 2.45\times10^{-3}M_{\rm ZAMS}^2 \nonumber \\
    &\quad -1.41\times10^{-2}M_{\rm ZAMS} + 2.94.
\end{align}

The $R_{\rm MAX}$ prescription described here is applicable exclusively to stars in their giant phase with initial masses below 100 $M_\odot$. Beyond this mass threshold, we assume a consistent behavior in the form of log$R_{\rm MAX}$($M_{\rm ZAMS}\geq$\,100\,$M_\odot$)\,=\,log$R_{\rm MAX}$(100\,$M_\odot$), which yields $R_{\rm MAX}$($M_{\rm ZAMS}\geq$\,100\,$M_\odot$)\,$\sim$\,955~$R_\odot$. In light of the negligible differences observed in $R_{\rm MAX}$ among low-metallicity models, we have decided in our {\tt StarTrack} framework to employ the $R_{\rm MAX}$($M_{\rm ZAMS}$, 0.1~$Z_\odot$) prescription for stars at any metallicity below 0.5~$Z_\odot$. For metallicities exceeding the 0.5~$Z_\odot$ threshold, we transition to the $R_{\rm MAX}$($M_{\rm ZAMS}$, $Z_\odot$) variant.

\subsection{Effect on stellar evolution}

\begin{figure*}[ht]
    \centering
        \includegraphics[width=0.465\textwidth]{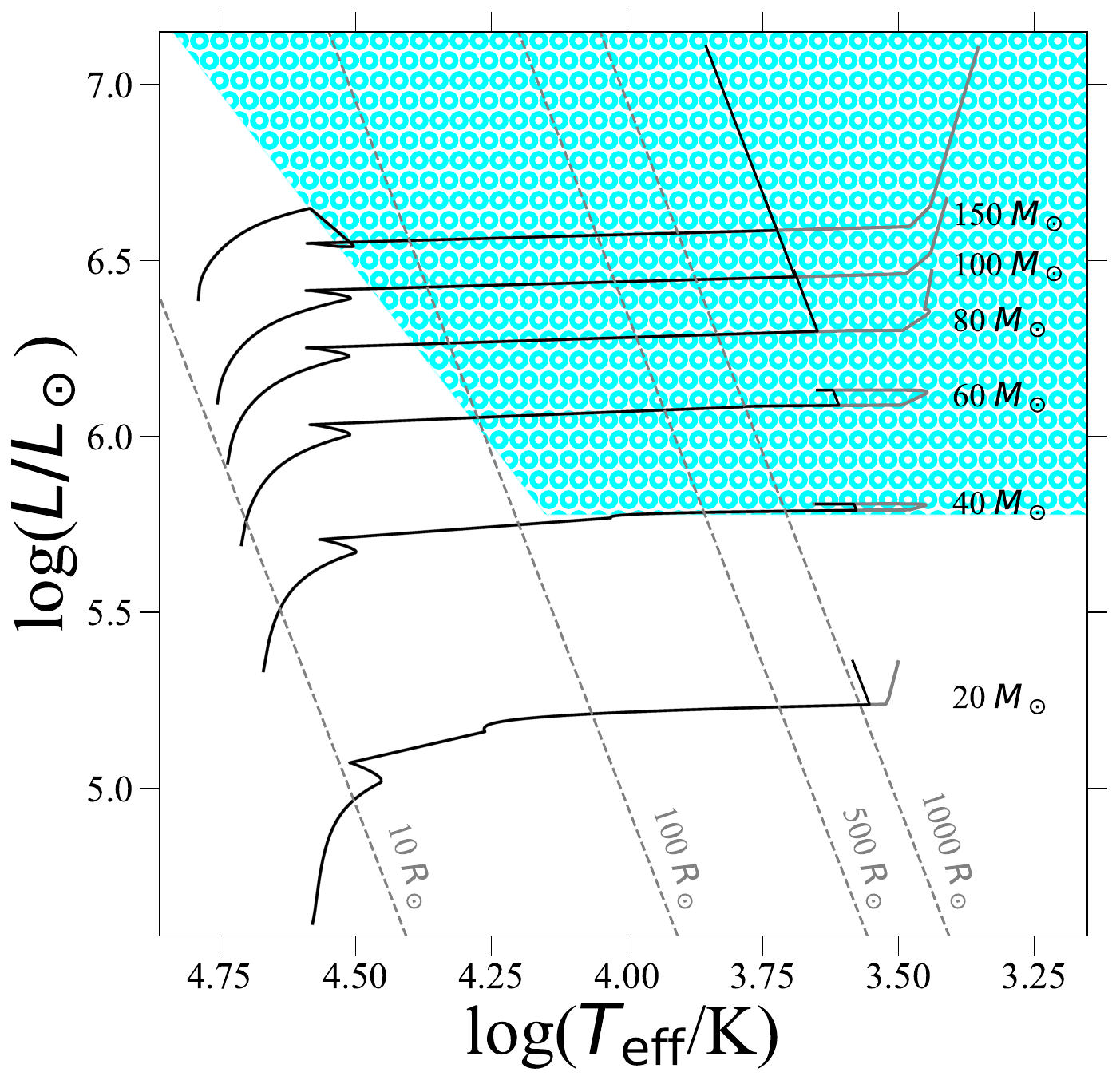}
        \centering
        \includegraphics[width=0.465\textwidth]{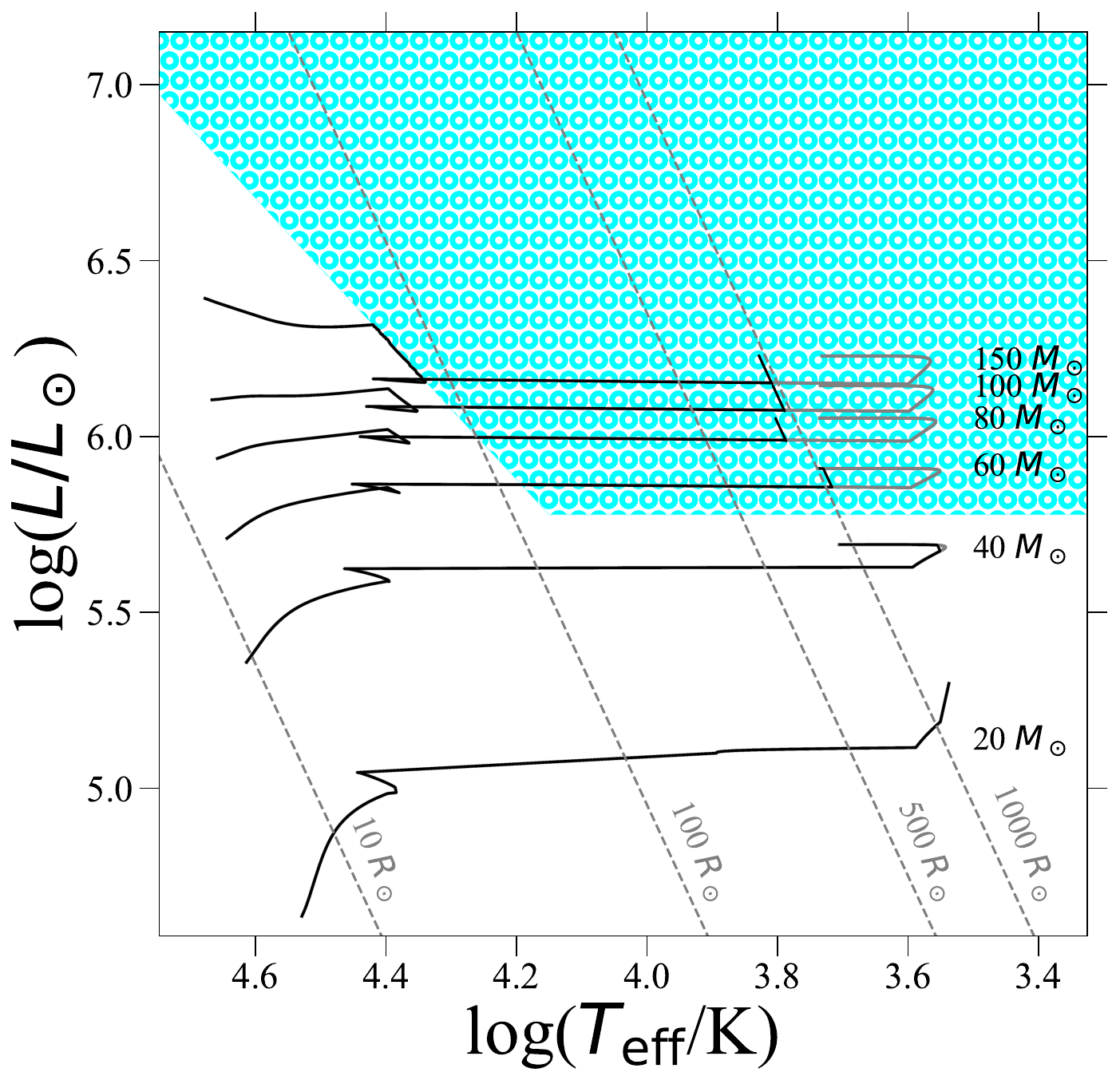}  %
    \caption{HR diagram for seven {\tt StarTrack} stellar tracks simulated with our $R_{\rm MAX}$ prescription for $Z$ = 0.5~$Z_\odot$ (left panel) and $Z =$~$Z_\odot$ (right panel). The initial masses are shown above their respective stellar tracks. The grey lines represent how the stars would evolve without limiting their expansion with our new radial prescription. The cyan area represents the area of the HR diagram beyond the HD limit. Despite the differences between M0 and M0$_{RMAX}$, in every case with $M_{\rm ZAMS}\gtrsim$40 $M_\odot$, the $R_{\rm MAX}$ limit to stellar expansion only takes place in our models when the stars are already in the LBV regime (which means that, according to our wind prescription, the wind mass loss does not change) or for relatively short periods of time that do not significantly impact the final properties of COs.
    }
    \label{fig:RMAX_HR}
\end{figure*}

In order to understand the consequences of using our $R_{\rm MAX}$ prescription on stellar evolution, we show in Figure \ref{fig:RMAX_HR} an HR diagram with seven {\tt StarTrack} evolutionary tracks at $Z$ = 0.00142 , 0.0142 and with $M_{\rm ZAMS}$ between 20 and 150 $M_\odot$. We simulated these stars with M0$_{\rm RMAX}$, which is noticeable from the tracks following their respective isoradii during the latest stages of their evolution. As a reference, we also show with the grey lines how these stellar tracks evolve during their latest stages without any $R_{\rm MAX}$ prescription (M0). The teal area is the part of the HR diagram that, according to the \cite{Hurley_2000} prescription, is beyond the HD limit:

\begin{equation}
      {\rm log}L > 6\times10^5
\label{eq:HDlimit_L}
\end{equation}
and
\begin{equation}
      10^{-5}\times R \times \sqrt{L} > 1 \hspace{0.2 cm} R_\odot L^{1/2}_\odot,
\label{eq:HDlimit_R}
\end{equation}

where $L$ stellar luminosity in $L_\odot$ and $R$ stellar radius in $R_\odot$. In our models a star crossing this boundary will enter its Luminous Blue Variable (LBV) phase, with a wind mass loss $\dot{M}_{\rm lbv}$ of

\begin{equation}
    \dot{M}_{\rm lbv}=f_{\text{LBV}} 10^{-4}[M_\odot\,{\rm yr}^{-1}]  \hspace{2.5 cm} f_{\rm lbv}=1.5
\end{equation}

This value is meant to account for both LBV wind mass loss and possible LBV shell ejections. $f_{\rm LBV}$ was originally calibrated in \cite{Belczynski_2010} to reproduce the most massive Galactic BHs from the catalogues of that time (e.g. \citealt{Orosz_2003,Casares_2007,Ziolkowski_2008}). Since our LBV mass loss prescription was meant as an average between the wind-driven mass loss and the super-Eddington mass ejection events, the mass lost of a star beyond the HD limit is not dependent in our models by its position on the HR diagram. Stars with $M_{\rm ZAMS}\gtrsim$\,40 $M_\odot$ for $Z$ = 0.0014 and  $M_{\rm ZAMS}\gtrsim$\,50 $M_\odot$ for $Z$ = 0.014 reach their maximum expansion when they are already in their LBV phase. This means that, at least for that mass range, for all intents and purposes the use of our $R_{\rm MAX}$ prescription does not affect stellar evolution in any way besides its intended effect on binary interactions. We nevertheless argue that if our $R_{\rm MAX}$ prescription was combined with different LBV winds models that do take into account the position of stars in the HR diagram (or even just the effective temperature), our $R_{\rm MAX}$ prescription would more considerably change the evolution of stars. We highlight that at high metallicities, the stellar tracks that do not cross the HD-limit almost do not even expand beyond their respective $R_{\rm MAX}$. For $Z$ = 0.0014, despite the track at $M_{\rm ZAMS}$\,=\,20\,$M_\odot$ does not cross the HD limit and therefore possesses temperature-dependent winds, the cumulative time at which M0 and M0$_{\rm RMAX}$ diverge in the HR diagram for these initial conditions is less than 0.2 Myr. During this time the two stars do lose different amounts of envelope mass, but after their collapse the BH from M0 will have a mass of 7.24 $M_\odot$, while the one from M0$_{\rm RMAX}$ will have one of 7.29 $M_\odot$, which is a negligible difference. This comes in agreement with the results presented in \cite{Romagnolo_2023}, where the use of our $R_{\rm MAX}$ prescription with {\tt StarTrack} does not alter the estimates of the distribution of the least massive BHs in BH-BH mergers within a redshift $z$\,<\,2.

\section{T$\dot{\rm Z}$O and QS lifetimes}
\label{sec:lifetimes}

\begin{figure}
\includegraphics[width=0.485\textwidth]{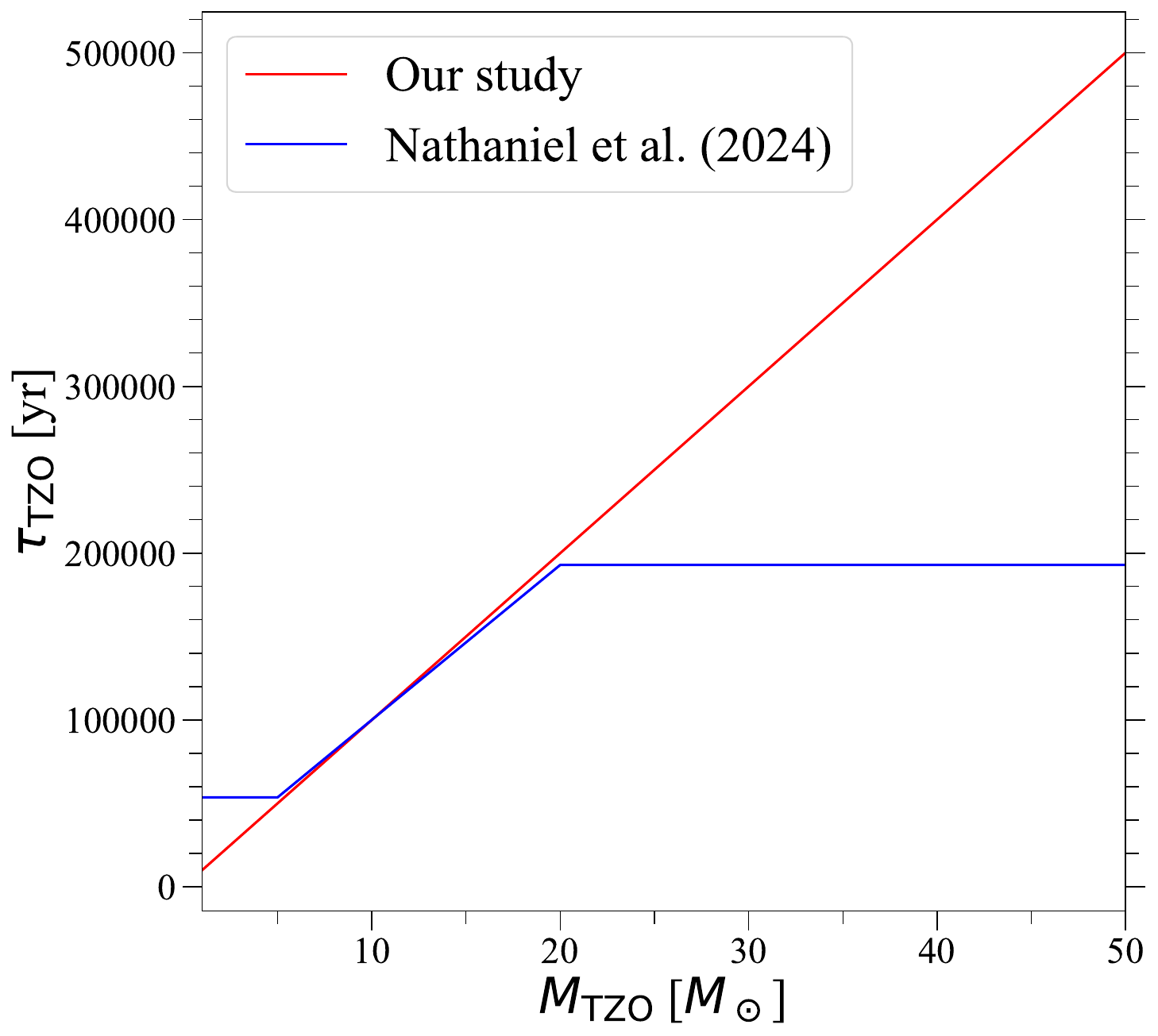}
\caption{
Comparison of the T$\dot{\rm Z}$O lifetime models between our work and \cite{Nathaniel_2024}
}
\label{fig:TZO_tau}
\end{figure}

We show in Figure~\ref{fig:TZO_tau} the different models for the lifetimes of T$\dot{\rm Z}$Os that our study and \cite{Nathaniel_2024} adopt up to $M_{\rm TZO}$\,=\,50~$M_\odot$. For objects below\,(above) 5\,(20)~$M_\odot$ our model predicts lifetimes that are lower\,(higher) than the ones reported in \cite{Nathaniel_2024}, while between 5 and 20~$M_\odot$ the values are similar.

\end{document}